\documentclass{aa} 
\usepackage[varg]{txfonts}
\usepackage{color}
\usepackage{longtable,lscape}
\usepackage{subcaption}

\newcommand{\kms} {km\,s$^{-1}$}

\newcommand{\Msun} {$\mbox{M}_{\sun}$}
\newcommand{\Mstar} {$\mbox{M}^{\star}$}

\newcommand{\oii}{[O\,{\sc ii}]$\,\lambda\lambda$3726,3729}
\newcommand{\ciii}{C\,{\sc iii}]$\,\lambda\lambda$1907,1909}
\newcommand{\lya}{Ly$\alpha$}

\begin{document}

\title{The MUSE Hubble Ultra Deep Field Survey\thanks{Based on observations made with ESO telescopes at the La Silla-Paranal Observatory under programmes 094.A-0289(B), 095.A-0010(A), 096.A-0045(A) and 096.A-0045(B).}:\\ IX. Evolution of galaxy merger fraction since $z\approx 6$}
\author{E.Ventou\inst{1}
\and T. Contini\inst{1}
\and N. Bouché\inst{1}
\and B. Epinat\inst{1,2}
\and J. Brinchmann\inst{3,4} 
\and R. Bacon\inst{5}
\and H. Inami\inst{5}
\and D. Lam\inst{3} 
\and A. Drake\inst{5}
\and T. Garel \inst{5}
\and L. Michel-Dansac \inst{5}
\and R. Pello \inst{1}
\and M. Steinmetz\inst{6}
\and P.M. Weilbacher \inst{6}
\and L. Wisotzki \inst{6}
\and M. Carollo\inst{7}
}
\institute{Institut de Recherche en Astrophysique et Planétologie (IRAP), Université de Toulouse, CNRS, UPS, F-31400 Toulouse, France
\and Aix Marseille Université, CNRS, LAM, Laboratoire d'Astrophysique de Marseille, UMR 7326, F-13388, Marseille, France
\and Leiden Observatory, Leiden University, P.O Box 9513, 2300 RA Leiden, The Netherlands
\and Instituto de Astrof{\'\i}sica e Ci{\^e}ncias do Espaço, Universidade do Porto, CAUP, Rua das Estrelas, PT4150-762 Porto, Portugal
\and Univ Lyon, Univ Lyon1, Ens de Lyon, CNRS, Centre de Recherche Astrophysique de Lyon UMR5574, F-69230, Saint-Genis-Laval, France
\and Leibniz-Institut f\"ur Astrophysik Potsdam (AIP), An der Sternwarte 16, 14482 Potsdam, Germany
\and Institute for Astronomy, Department of Physics, ETH Zurich, 8093 Zurich, Switzerland
}

\abstract
{We provide, for the first time, robust observational constraints on the galaxy major merger fraction up to $z\approx 6$ using spectroscopic close pair counts.  Deep Multi Unit Spectroscopic Explorer (MUSE) observations in the Hubble Ultra Deep Field (HUDF) and Hubble Deep Field South (HDF-S) are used to identify 113 secure close pairs of galaxies among a parent sample of 1801 galaxies spread over a large redshift range ($0.2<z<6$) and stellar masses ($10^7-10^{11}$\Msun), thus probing about 12 Gyr of galaxy evolution. Stellar masses are estimated from spectral energy distribution (SED) fitting over the extensive UV-to-NIR HST photometry available in these deep Hubble fields, adding Spitzer IRAC bands to better constrain masses for high-redshift ($z\geqslant 3$) galaxies. These stellar masses are used to isolate a sample of 54 major close pairs with a galaxy mass ratio limit of 1:6.  Among this sample, 23 pairs are identified at  high redshift ($z\geqslant 3$) through their \lya\ emission. 
The sample of major close pairs is divided into five redshift intervals in order to probe the evolution of the merger fraction with cosmic time. Our estimates are in very good agreement with previous close pair counts with a constant increase of the merger fraction up to $z\approx 3$ where it reaches a maximum of 20\%. At higher redshift, we show that the fraction slowly decreases down to about 10\% at $z\approx6$. The sample is further divided into two ranges of stellar masses using either a constant separation limit of $10^{9.5}$\Msun\ or the median value of stellar mass computed in each redshift bin. Overall, the major close pair fraction for low-mass and massive galaxies follows the same trend. 
These new, homogeneous, and robust estimates of the major merger fraction since $z\approx6$ are in good agreement with recent predictions of cosmological numerical simulations.}
\maketitle

\section{Introduction}
\label{sec:intro}
Galaxy mergers play a key role in the formation and evolution of galaxies (e.g.\, Baugh 2006; Conselice 2014), especially in a $\Lambda$CDM cosmology where structures of dark matter halos (DMH) grow hierarchically (e.g.\, White \& Rees 1978).
These events have an important impact on the evolution of galaxies, such as their mass assembly (De Lucia \& Blaizot 2007; Guo \& White 2008; Genel et al. 2009; Rodriguez-Gomez et al. 2016; Qu et al. 2017), and their star formation history (Mihos \& Hernquist 1996; Somerville et al. 2001). Mergers are also responsible for drastic changes in galaxy morphologies, internal structures, and dynamics (e.g.\, Mihos \& Hernquist 1994; Naab \& Burkert 2003; Bell et al.\,2008; Perret et al.\, 2014; Lagos et al.\,2017). Understanding the role of mergers in the evolution of galaxies and their importance relative to other processes, such as cold gas accretion (e.g. Keres et al. 2005; Ocvirk et al. 2008; Genel et al. 2008), is thus a key aspect of galaxy formation models.

The most simple and direct way to investigate the role of mergers in galaxy evolution is to count the number of observed events. There are several approaches for the identification of mergers in the universe. The occurrence of morphologically disturbed systems, through visual inspection (e.g.\, Brinchmann et al. 1998; Bundy et al. 2005; Kampczyk et al. 2007) or quantitative measurements (e.g.\, Abraham et al. 1996; Conselice et al. 2000, 2003, 2009; Lotz et al. 2008; Lopez-Sanjuan 2009a,b; Casteels et al. 2014), has been widely used thanks to deep and high-resolution images such as those from HST. A second approach is to count close pairs of galaxies, i.e.\, two galaxies with low values of projected angular separations ($\leqslant 25$ h$^{-1}$kpc) and line-of-sight relative radial velocities ($\leqslant 500$ \kms). Simulations have shown that the vast majority of pairs meeting these criteria indeed merge on reasonable timescales, typically shorter than 1 Gyr (Boylan-Kolchin et al. 2008; Kitzbichler \& White 2008; Jian et al. 2012; Moreno et al 2013). However, these different methods of selecting merger candidates might be sensitive to different stages in the merging process, for example\, pre-merging or early merging for close pair counts and ongoing merging or post-merging from morphological identification. Observational constraints on the merger fractions can then differ by up to an order of magnitude and yield very different redshift evolution depending on the method adopted (see next paragraphs). 

Major close pairs, usually defined to be those involving galaxies with a mass ratio greater than 1:4, are now well studied up to $z\sim 1$. The early measurements using photometric redshifts (Patton et al. 1997; Le Fèvre et al. 2000) have been superseded by spectroscopic surveys, confirming physical pairs from the redshift measurement of both components (e.g.\, Lin et al. 2008; de Ravel et al. 2009; Lopez-Sanjuan et al. 2012, 2013; Tasca et al. 2014), even if some recent photometric surveys, such as ALHAMBRA or SHARDS, allow the computation of accurate close pair fractions (Ferreras et al. 2014; Lopez-Sanjuan et al. 2015). 

In the nearby universe, the major merger fraction is only about $2\%$  (e.g.\, Patton \& Atfield 2008; Casteels et al. 2014). But this fraction increases significantly up to  $z \sim 1$ indicating that major mergers could be responsible for $20\%$ of the growth of stellar mass density of galaxies from $z \sim 1$ (e.g.\, Bundy et al. 2009; de Ravel et al. 2009; Lopez-Sanjuan et al. 2012). The evolution of the major merger fraction as a function of redshift is commonly parameterized as a power law of the form $(1+z)^{m}$. Even if the pair fraction is thought to be an increasing function of redshift, the range of reported values is almost unconstrained with $m=0-5$ (e.g.\, Le Fevre et al. 2000; Lin et al. 2004, 2008; Kampczyk et al. 2007; Kartaltepe et al. 2007; de Ravel et al. 2009; Lotz et al. 2011, Keenan et al. 2014). However, these discrepancies could be decreased when introducing an observability timescale for identifying galaxy mergers (Lotz et al. 2011). 

Beyond $z\sim 1$, direct measurements of the major merger fraction are still limited. Previous attempts to measure the major merger rate at $z > 1$ have focussed on the identification of merger remnants from morphological studies (e.g.\, Conselice et al. 2008, 2011; Bluck et al. 2012) or photometric close pairs (e.g.\, Ryan et al. 2008; Bluck et al. 2009; Man et al. 2012, 2016). These studies find an increase of the merger rate up to $z\sim 2-3$ but with a large scatter between different measurements.  Estimates of major merger rates from spectroscopic close pairs, which is a much more robust way to confirm the physical closeness of the two galaxies, are still sparse with a handful of merger systems identified in Lyman-break galaxy samples (Cooke et al. 2010),  MASSIV (Lopez-Sanjuan et al. 2013) and VVDS/VUDS surveys (Tasca et al. 2014). These studies converge towards a fraction around $20\%$ at these redshifts. Because of the difficulty of detecting close spectroscopic pairs of galaxies, no measurements beyond $z\sim 3$ have been reported so far. 

The fact that the fraction of major mergers remains constant or turns over beyond $z\sim 1$ is in agreement with the prediction of recent cosmological simulations, such as Horizon-AGN (Kaviraj et al. 2015), EAGLE (Qu et al. 2017) and Illustris (Snyder et al. 2017).
It remains also an intriguing question, down to which galaxy masses mergers will play an important role. There are indications in the nearby universe that low-mass dwarf galaxies experienced strong gravitational interactions and/or merging events in the past (e.g.\,Harris \& Zaritsky 2009; Besla et al. 2012; Koch et al. 2015).  But estimates on the major merger rate in the distant universe have been restricted so far to massive galaxies alone ($\geqslant 10^{10}$\Msun). 

This paper aims to provide new constraints on the evolution of the galaxy major merger fraction over the last 12 billion years, i.e.\, extending up to redshift $z\sim 6$, and over a large range of galaxy masses.
This analysis is based on deep MUSE observations in two fields: one in the Hubble Deep Field South (HDF-S) and one in the Hubble Ultra Deep Field (HUDF). Thanks to its wide field of view and unprecedented sensitivity, MUSE enables us to perform deep spectroscopic  surveys without any pre-selection of galaxies, which was the main drawback of previous spectroscopic surveys. This new and powerful instrument is thus perfectly suited to identify close pairs of galaxies at very high redshift ($z > 3$) with spectroscopic redshifts, and to probe a much larger range of stellar masses than before. As we are exploring new territories with MUSE, the conversion of the merger fractions into merger rates is postponed to a second paper. Indeed, the merger (or pair observability) timescale, usually derived from the prescription of Kitzbichler \& White (2008), is a model-dependent parameter. which is so far unconstrained for very high-redshift and/or low-mass galaxies (see e.g.\, Snyder et al.\,2017). 

This paper is organized as follows. In section\,\ref{sec:data}, we introduce the MUSE data sets used to detect galaxy close pairs. We describe the method to identify close pairs in the spectroscopic redshift catalogues, how we can recover the systemic redshift of Ly$\alpha$ emitters, and the main limitations of the method in Sect. 3. We make the distinction between minor and major close pairs according to the stellar mass ratio between the two galaxies in section\,\ref{sect:masses}. We give an estimate of the major merger fraction evolution up to $z\sim 6$ and compare our results with recent numerical simulations in section\,\ref{sec:fraction}. 
A summary and conclusion are given in section\,\ref{sec:conclusion}.

Throughout our analysis, we use a standard $\Lambda$CDM cosmology with  $H_0=100 h$ kms$^{-1}$ Mpc$^{-1}$, $h=0.7$, $\Omega_{m}= 0.3$, $ \Omega_{\Lambda}= 0.7$.

\section{MUSE data set}
\label{sec:data}
This analysis is based on MUSE observations in the Hubble Deep Field South (HDF-S; Williams et al., 2000) and the Hubble Ultra Deep Field (HUDF; Beckwith et al., 2006).
MUSE field of view covers a $1\times 1$ arcmin$^{2}$ area over a wavelength range of $4750-9300$\AA. 

\subsection{Hubble Deep Field South}
The HDF-S was observed during a MUSE commissioning run in August 2014, resulting in a single field of 27 hours of total exposure time centred around $\alpha=22h 32' 55.64"$ and $\delta=-60^o 33' 47"$. The data cube contains 
spectra with a spectral resolution of $\sim 2.3$ \AA\ and a spatial resolution ranging between $0.6$\arcsec\ for the red end of the spectral range and $0.7$\arcsec\ in the blue. The spectroscopic redshift of 189 sources were accurately measured up to a magnitude of $I_{814}=29.5$.  Details on the data reduction, source identification, redshift determination, and source catalogue can be found in Bacon et al.\,(2015).
 
 \subsection{Ultra Deep Field-Mosaic}
The HUDF region was observed with MUSE during Guaranteed Time Observations from September 2014 to February 2016, resulting in one medium-deep mosaic of nine MUSE pointings covering the entire HUDF and one single MUSE deep ($\sim 31$h) pointing, \textsf{udf-10} (see below). The UDF-Mosaic consists of nine MUSE fields of $1\times 1$ arcmin$^{2}$, which resulted in a field of $3.15 \times 3.15$ arcmin$^{2}$ with an average of $10$ hours exposure time. The achieved spatial resolution is $0.71$\arcsec\ (at $4750$ \AA) and $0.57$\arcsec\ (at $9350$ \AA), and the spectral resolution ranges from $3.0$ \AA\ at the blue end to $2.4 $ \AA\ at $7500$ \AA\ (see Bacon et al.\,2017 for more details). Overall the spectroscopic redshifts of 1439 sources were measured (Inami et al.\,2017).
 
\subsection{\textsf{Ultra Deep Field-10}}
\label{udf10}
With $31$ hours of exposure time, which consist of $21$ hours of \textsf{udf-10} pointing and $10$ hours of Mosaic pointing, \textsf{udf-10} is the deepest field observed with MUSE up to now (Bacon et al.\,2017). This $1.15$ arcmin$^{2}$ field is located in the XDF area,  centred around $\alpha=03h 32' 38.7"$ and $\delta=-27^o 46' 44"$ and overlapping with the UDF-Mosaic. The spectral and spatial resolution are similar to those for the UDF-Mosaic.  In this region, 313  spectroscopic redshifts were measured (Inami et al.\,2017). To avoid confusion, from now on, the UDF-Mosaic that we used for this analysis corresponds to the whole Mosaic field without its \textsf{udf-10} region. For this overlapping region we used the 31 hours \textsf{udf-10} data. 

\section{Detection of close galaxy pairs}
\label{sec:detection}

\subsection{Parent galaxy sample}
\label{parent}

 \begin{figure}[t]
        \includegraphics[width=\columnwidth]{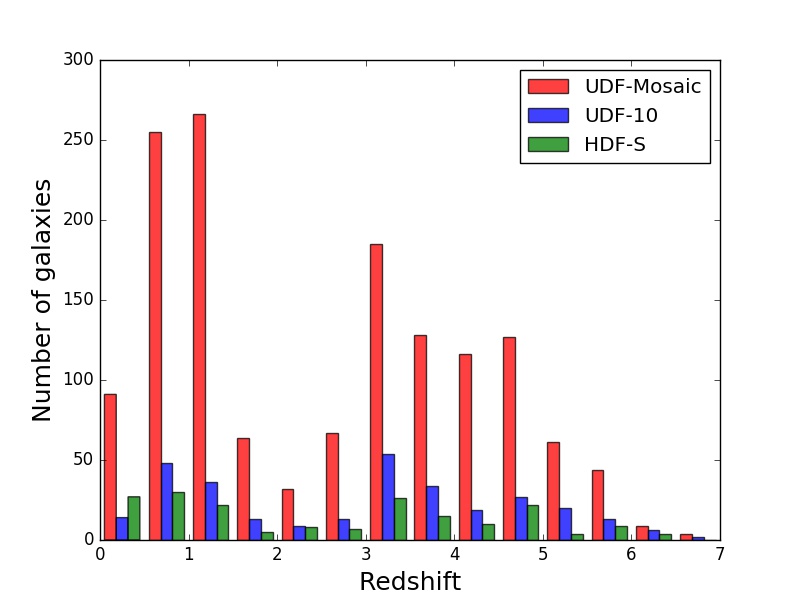}
   \caption{Spectroscopic redshift distribution of galaxies in the three MUSE data cubes used in this analysis.}
        \label{fig:zdistrib}
\end{figure}

The parent sample used for this analysis includes all galaxies with measured spectroscopic redshift from the catalogues associated with each of the three fields: HDF-S, \textsf{udf-10} and UDF-Mosaic (for more details see Inami et al. 2017 and Bacon et al. 2015). As explained in sect.\,\ref{udf10}, we removed all sources present in the \textsf{udf-10} region from the UDF-Mosaic catalogue.  

The combined fields result in a parent sample of 1801 galaxies with spectroscopic redshift assigned with a confidence level from 3 to 1. A confidence flag of 3 or 2 means that the redshift is secure, with a measurement based on multiple features or a clearly identified single feature (\oii\ or \ciii\ doublet, asymmetric \lya\ line). For the lowest confidence level of 1, the redshift was determined by a single feature but with uncertainties on the nature of this feature (no clear doublet or asymmetry).The global estimate of the redshift uncertainty  corresponds to $\sigma_{z}=0.00012(1+z)$ (Inami et al.\, 2017). Figure\,\ref{fig:zdistrib} shows that our parent sample extends over a broad range of spectroscopic redshifts, extending up to $z\approx 7$. Compared to HDF-S and \textsf{udf-10} redshift distributions, the histogram in UDF-Mosaic peaks at $z\approx 1$ because of an over-dense structure detected around this redshift. Between $1.5\leqslant z\leqslant 2.8$, the interval described as the redshift desert for optical surveys, there is a dearth of spectroscopic measurements because the instruments we used are sensitive to strong emission-line galaxies up to $z=1.5$ with \oii\  and above $z\geqslant 2.8$ with \lya, but our results are missing such bright emission lines in between. Thereby the sources detected in this range tend to be continuum-bright galaxies corresponding to a more massive galaxy population (see sect.\,\ref{sec:zbins}). Their redshifts are based on absorption features or CIII] emission.
 
 \begin{figure*}[t]
  \centering 
    \begin{tabular}{cc}
       \includegraphics[width=0.37\textwidth]{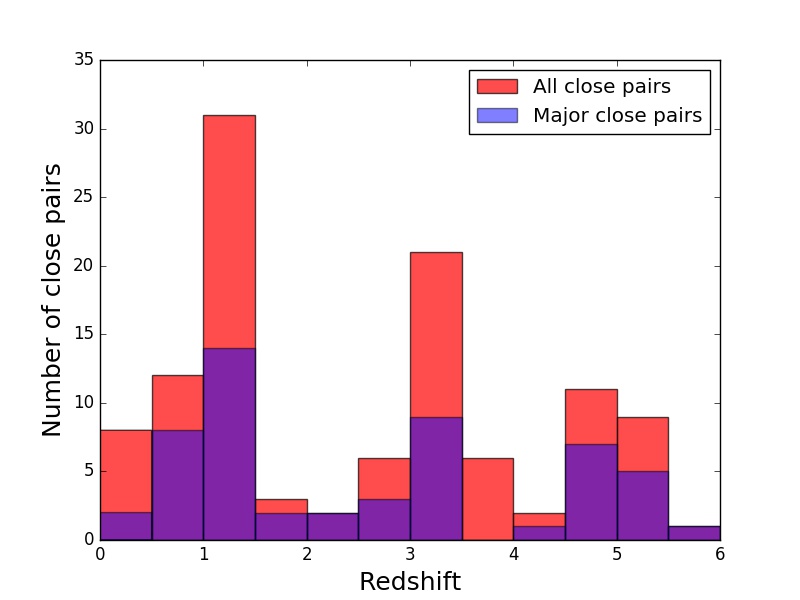}
 &
        \includegraphics[width=0.37\textwidth]{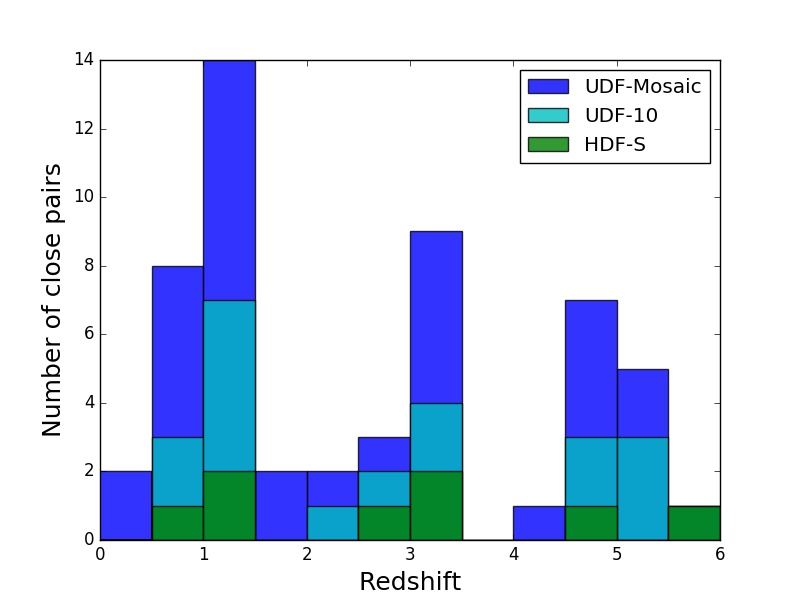} \\
        \end{tabular}
           \caption{{\it Left}: Redshift distribution of all the  galaxy  close pairs (red) and the contribution of major close pairs (purple). {\it Right}: Redshift histogram of the major close pairs showing the contribution of the different MUSE fields: UDF-Mosaic (dark blue), \textsf{udf-10} (light blue), and HDF-S (green).}
        \label{fig:z_histo}
   \end{figure*}

\subsection{Selection criteria for close pair}

We identified a close pair as a system of two galaxies within a limited projected separation distance in the sky plane, $r_p^{min}\leqslant r_p \leqslant r_p^{max}$, and a rest-frame relative velocity, $\Delta v \leqslant \Delta v_{max}$.
These parameters are computed as follows:
\begin{equation}
r_p=\theta \times d_A (z_m)
\end{equation}
 where $\theta$ is the angular  distance (in arcsec) between the two galaxies, $d_A (z_m)$ is the angular scale (in kpc arcsec$^{-1}$), and $z_m$ is the mean redshift of the two galaxies.
The rest-frame velocity is written as
\begin{equation}
\Delta v = \frac{c \times |z_1-z_2|}{(1+z_m)}
,\end{equation}
where $z_1$ and $z_2$ are the redshifts of each galaxy in the pair.

Previous observational  and  theoretical  studies revealed $25$h$^{-1}$kpc to be the 
approximate scale on which the majority of the  pairs start to exhibit interacting features such as tidal tails, bridges, distortions, or  enhancement of the star formation rate in the galaxies (Patton et al.\,2000; Alonso et al.\,2004; Nikolic et al.\,2004). We thus selected a limit of  $r_p^{max}=25$h$^{-1}$ kpc to select close pairs with a high probability of merging. For the  maximum rest-frame  velocity  difference  of  a  galaxy  pair,  $\Delta v_{max} = 500$ kms$^{-1}$ offers a good compromise between contamination and statistics. A  smaller  velocity  separation would reduce the sample size, which  limits  the  robustness of the pair statistics. These effects have also been discussed in Patton et al. (2000).

\subsection{Selection method}

From the spectroscopic parent sample of 1801 galaxies (see sect.\,\ref{parent}), we searched for close kinematic galaxy pairs following the projected separation distance and the rest-frame relative velocity criteria defined above. In order to assess the reliability of these pairs, we then extracted a sub-cube of approximately $60$ h$^{-1}$kpc around the position of the galaxy and created  narrowband images for each emission lines identified in the spectrum of the primary galaxy, which corresponds to the most massive galaxy in the pair. This procedure was found to be very helpful in constructing the final version of the spectroscopic catalogues (Bacon et al. 2015; Inami et al. 2017) by identifying and rejecting some spurious pairs (see sect.\,\ref{sec:limitations}). Finally, all the close pairs selected from the redshift catalogues and used in this analysis were checked and validated.

\subsubsection{Recovering the systemic redshift of \lya\ emitters}
For redshifts below $z\approx 2.8$, emission lines such as \oii\ and \ciii\  accurately trace the systemic redshift of the observed galaxy. However most spectroscopic redshifts for galaxies above $z\approx 2.8$ are derived from the peak of \lya\ emission line, which introduces uncertainties in redshift estimates since \lya\ is usually red-shifted by several hundreds of kms$^{-1}$ from systemic redshift (e.g. McLinden et al. 2011; Hashimoto et al. 2013; Erb et al. 2014; Shibuya et al. 2014). This could have a major impact on our pair selection at high redshift as this velocity shift is of the same order as the velocity criteria used to define a close pair. We must then find a way to correct the spectroscopic redshift of our \lya\ emitters before performing the selection of close pairs above $z\approx 2.8$.

Idealized models of radiative transfer (e.g.\, Verhamme et al. 2015) have predicted that the full width at half maximum (FWHM) of Ly$\alpha$ is correlated with the column density of the scattering medium, as is the velocity shift of the emission peak relative to the systemic velocity. 
This trend has been investigated recently to build an empirical relation between these two parameters (Verhamme et al., in prep.). This study includes a sample of \lya\ emitters from the UDF-Mosaic and \textsf{udf-10} in their data sets to investigate this relation. The  observed Ly$\alpha$ FWHM is thus used as a proxy to correct our Ly$\alpha$-based redshifts for this velocity offset.

We applied this correction to our parent sample using equation 2 of  Verhamme et al.\, (in prep.), and then performed our selection of close pairs with the corrected spectroscopic redshifts.
Although this correction impacts the ``true'' velocity difference between the galaxies in the pairs, it has a small impact on the final number of close pairs, with a variation of only three pairs, corresponding to $\approx 3$\% of the total number of pairs.

\subsubsection{Some limitations of the method}
\label{sec:limitations}

Because of the limited spatial resolution of MUSE data, it is nearly impossible to distinguish two galaxies within an angular separation of  $\theta \leqslant 0.7\arcsec$, which corresponds to an inner projected separation radius of $r_p^{min}\sim 3 - 5$ h$^{-1}$kpc  depending on the redshift. For most of these cases, galaxies are undergoing a merging process. These missing pairs are taken into account later (see sect.\,\ref{sec:fraction}) in the expression of the merger fraction.

In some cases, primary galaxies have a strong extended emission line that contaminates the spectrum of close companions, and as such, were detected as a close pair. Only a careful check in the data cube, for example by producing narrowband images around the line of interest, allowed us to separate these candidates from real spectroscopic pairs. This careful cleaning was applied iteratively on the incremented versions of the catalogue to reach a maximum of purity.

Since most of the spectroscopic redshifts are based on emission lines, we introduced a bias towards star-forming or active galaxies in the sample; thus, we are missing a significant percentage of continuum-faint quiescent galaxies. 

Finally, for close pairs with at least one galaxy with a low-confidence redshift (see sect.\,\ref{parent}), leading to ``unsecure'' pairs, we applied a lower weight than for secure pairs in the expression of the merger fraction (see sect.\,\ref{sec:fraction}). 

\subsection{Results}

Based on the method described above, we identified a total of 113 close pairs: 65 in the UDF-Mosaic, 31 in the \textsf{udf-10}, and 17 in the HDF-S, distributed over a broad range of redshifts, from $z\sim 0.2$ to $6$ (see Fig.\,\ref{fig:z_histo}, left panel).

\begin{figure*}[!h]
 \begin{subfigure}{1\textwidth}
 \centering
 \includegraphics[width=0.8\linewidth]{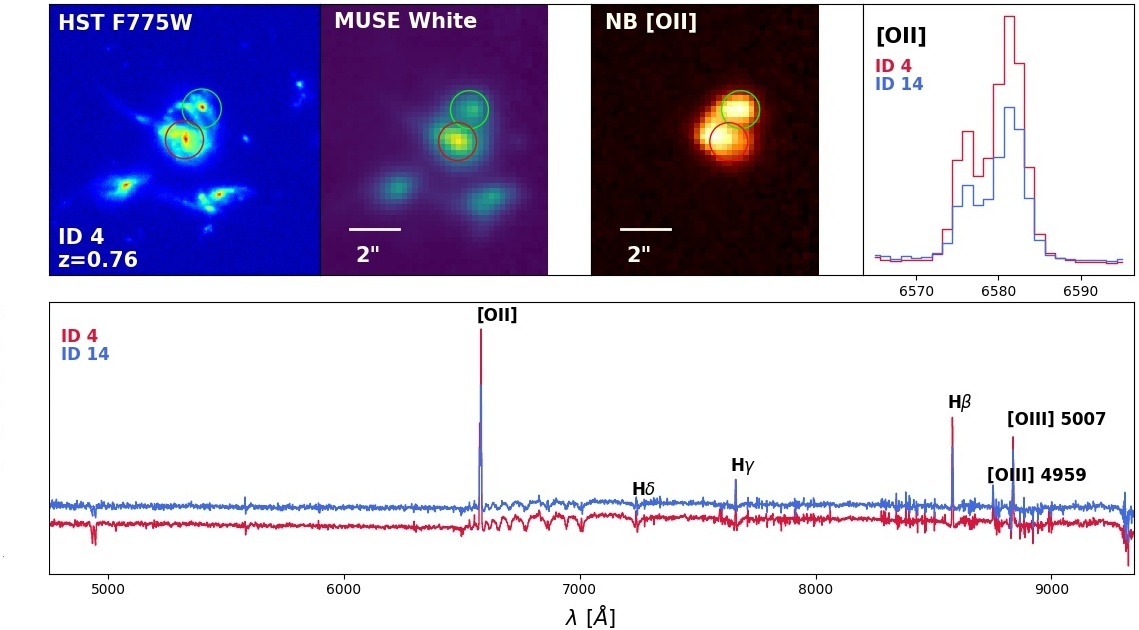}
 \caption{ }
 \label{p1}
 \end{subfigure}
 \begin{subfigure}{1\textwidth}
 \centering
 \includegraphics[width=0.8\linewidth]{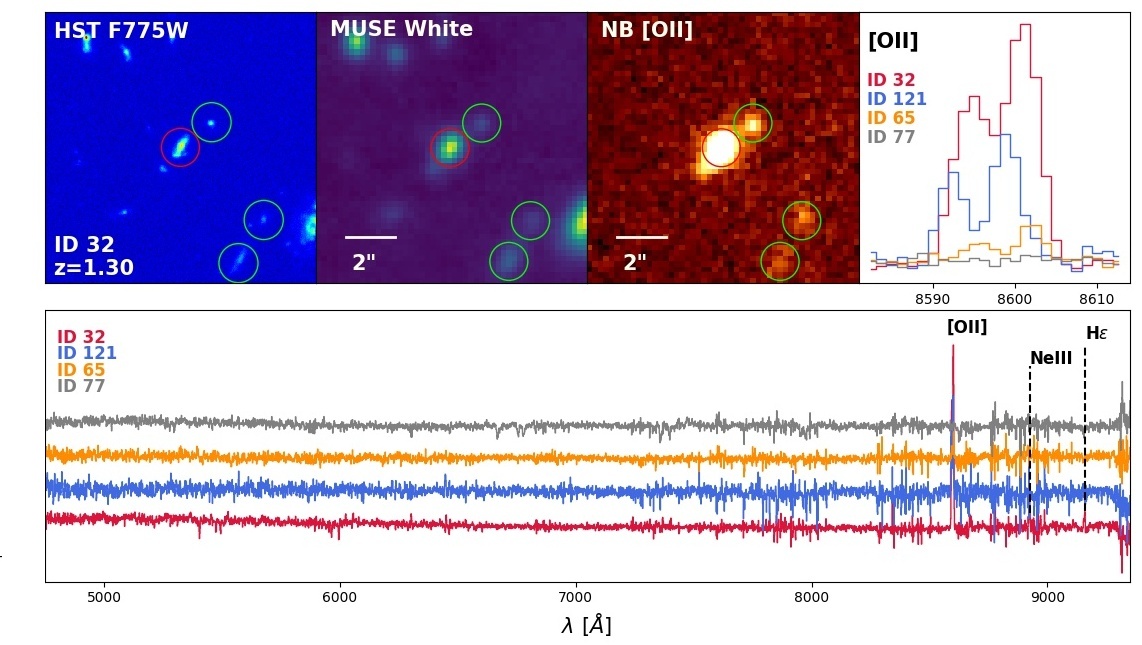}
 \caption{}
 \label{p2}
 \end{subfigure} 
   \caption{Examples of galaxy close pairs. {\it Top line, from left to right}: HST image in the F775W filter with the labelled MUSE ID and redshift of the primary galaxy, MUSE reconstructed white light image, narrowband image of one of the brightest emission lines of the pair, and the zoomed spectra around this line. Images are $10$\arcsec\ in linear size and centred around the primary galaxy, i.e.\, the most massive one, circled in red. The green circle(s) denote its companion(s). {\it Bottom line}: Spectrum (red for the primary and blue or other colours for its companion) over the whole wavelength range observed with MUSE, differentiated by an arbitrary offset. Fluxes are in arbitrary units. The main emission(absorption) lines are labelled in black(grey).\newline 
(a) A low redshift close pair of galaxies in \textsf{udf-10} at $z= 0.76$ with  $r_\mathrm{p}\sim 6  \ \mathrm{kpc}$ and  $\Delta v\sim 7$ \kms. This pair has a strong \oii\ emission line slightly of-centred, and shows signs of interactions such as tidal tails.\newline
(b) A quadruplet of galaxies in \textsf{udf-10} at $z=1.30$ within a projected separation distance of $r_\mathrm{p} \sim 41  \ \mathrm{kpc} $ between the primary galaxy, MUSE ID 32, in the centre, and the most distant satellite galaxy at the bottom right of the image and within a maximum rest-frame velocity of $\Delta v\sim 220$ \kms. The MUSE ID of the companion galaxies are, from top to bottom, 121, 77, and 65. Objects 32, 121, and 65 all have a secure spectroscopic redshift with a confidence flag in the measurement of 3, whereas object 77 has a confidence level of 1, which is taken into account in the computation of the fraction (see \ref{sec:frac_evol}). The 1D spectrum of this galaxy shows a much fainter \oii\ emission than the other galaxies, but the galaxy is clearly identified in the narrowband image. The strong absorption lines in its spectrum belong to another source, ID 18. In such a case of multiple close pairs, where some of the paired galaxies have another partner, the number of close pairs corresponds to the number of satellite galaxies, i.e.\, we account for 3 close pairs in this system.}
 \label{im_pair1}
\end{figure*}

\begin{figure*}[h]
 \begin{subfigure}{1\textwidth}
 \centering
 \includegraphics[width=0.8\linewidth]{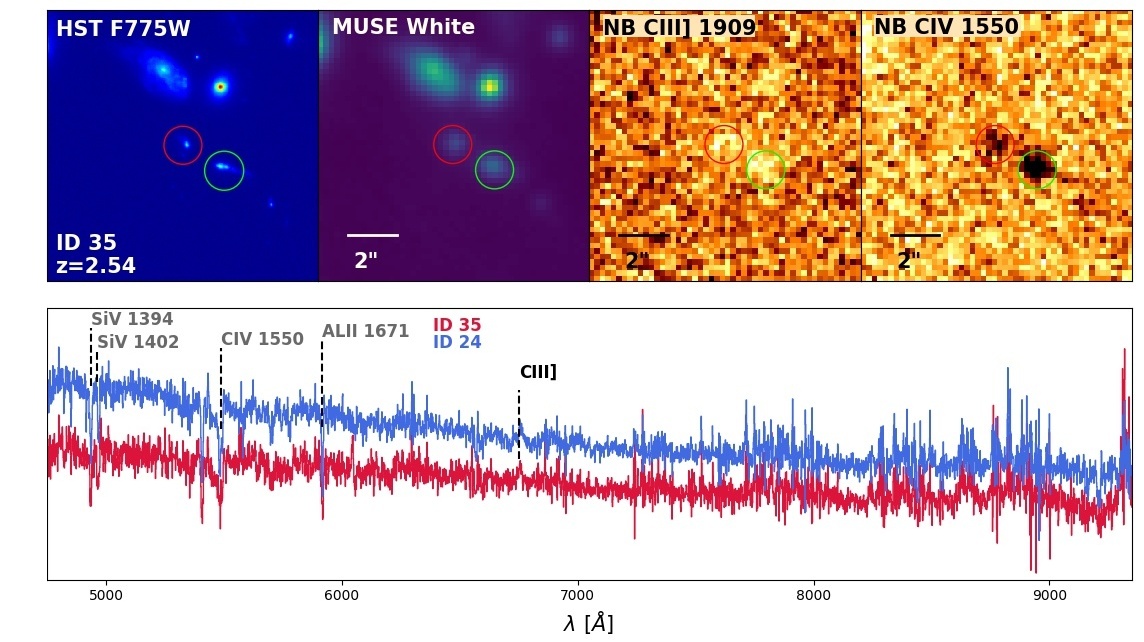}
 \caption{ }
 \label{p3}
 \end{subfigure} 
  \begin{subfigure}{1\textwidth}
 \centering
 \includegraphics[width=0.8\linewidth]{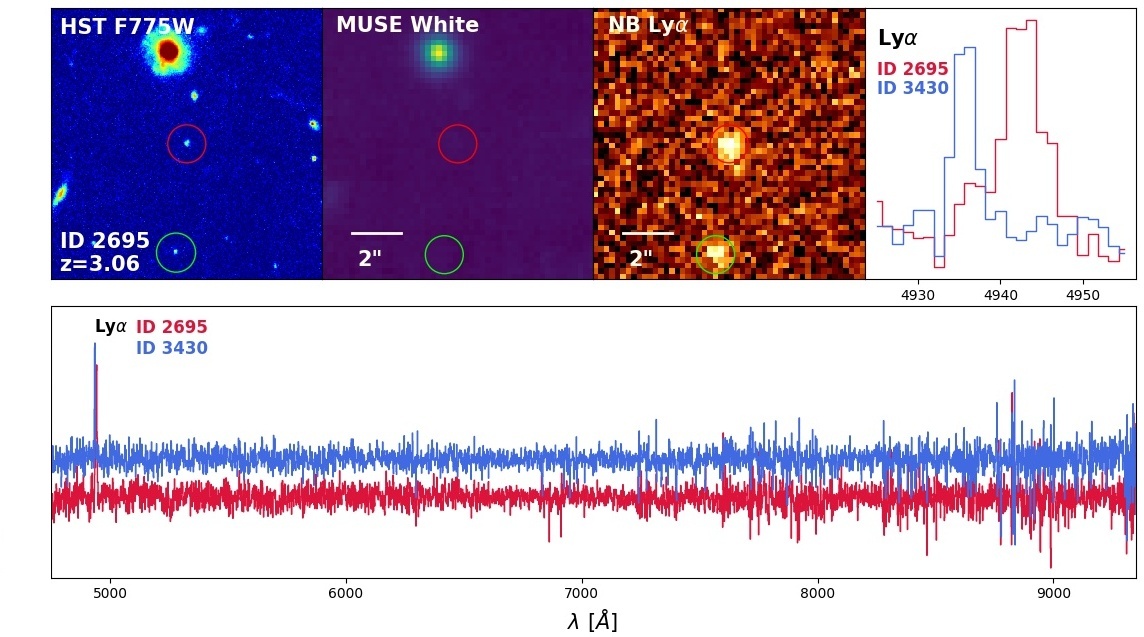}
 \caption{}
 \label{p4}
 \end{subfigure}
  \caption{Same as Figure \ref{im_pair1}.\newline
  (a) A close pair in \textsf{udf-10} at a redshift of $z=2.54$ with $r_\mathrm{p}\sim 15  \ \mathrm{kpc}$ and $\Delta v\sim 6$ \kms. This is a good example of the galaxy pair population detected in the redshift desert bin. The two continuum-bright galaxies reveal a faint \ciii\ emission line, as is shown in the first narrowband image, but are clearly identified thanks to their strong absorption lines.\newline(b) A close pair of \lya\ emitters (LAE) in the UDF-Mosaic at $z=3.06$, one of the three close pairs with a rest-frame relative velocity higher than $300$ \kms\ with $\Delta v\sim 317$ \kms\ and $r_\mathrm{p}\sim 31 \ \mathrm{kpc}$.}
 \label{im_pairs2}
\end{figure*}

\begin{figure*}[h]

 \begin{subfigure}{1\textwidth}
 \centering
 \includegraphics[width=0.8\linewidth]{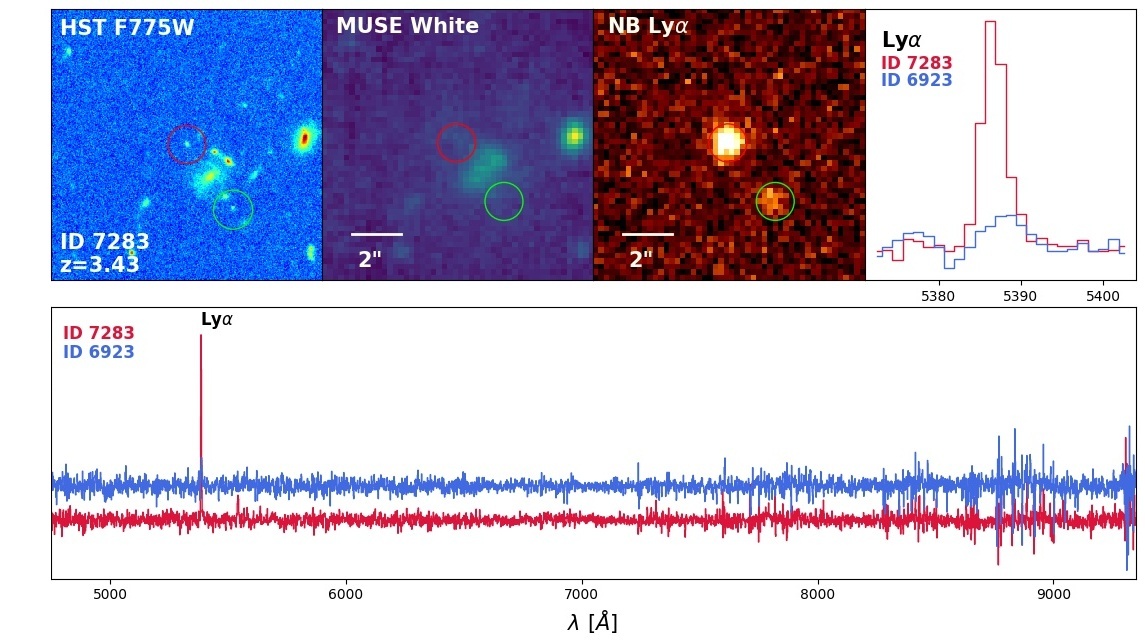}
 \caption{}
 \label{p5}
 \end{subfigure} 
 
 \begin{subfigure}{1\textwidth}
 \centering
 \includegraphics[width=0.8\linewidth]{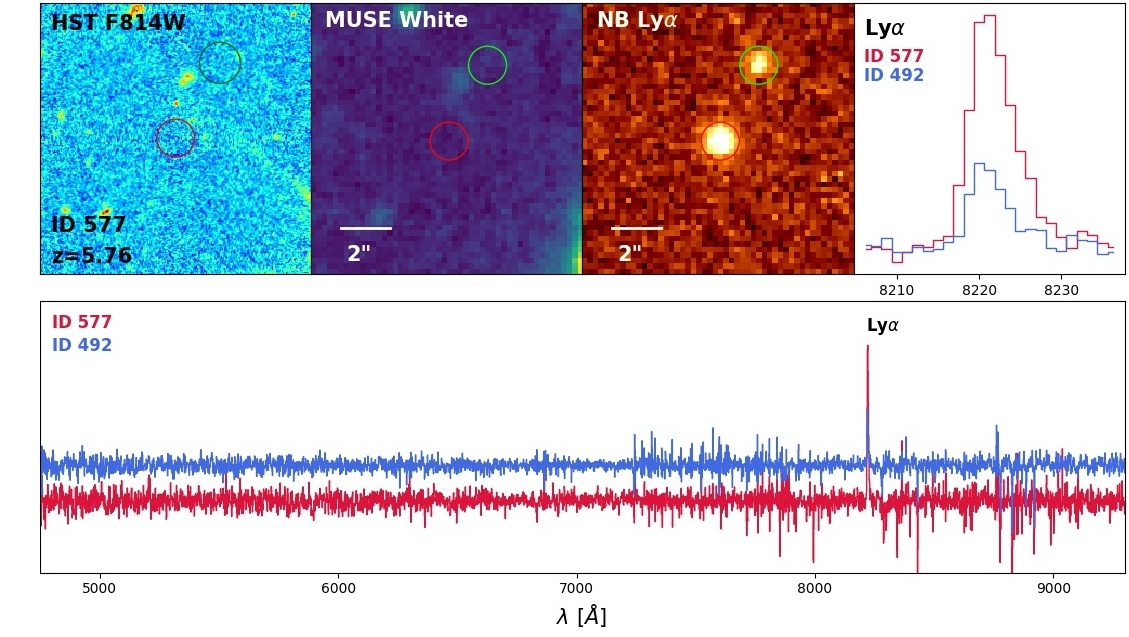}
 \caption{}
 \label{p6}
 \end{subfigure} 
 \caption{Same as Figure \ref{im_pair1}.\newline
 (a) A close pair at $z=3.43$ in the UDF-Mosaic with the primary galaxy showing a strong \lya\ emission, compared to its companion, which has a much fainter \lya\ emission. The two galaxies are separated with a projected distance of $r_\mathrm{p}\sim 22  \ \mathrm{kpc}$ and a difference in velocity of $\Delta v\sim 49$ \kms.\newline
 (b) At $z=5.76$, this close pair of LAE is the highest redshift pair of our sample, located in the HDF-S with $r_\mathrm{p}\sim 19  \ \mathrm{kpc}$ and $\Delta v\sim 16$ \kms.}
 \label{im_pairs3}
\end{figure*}

We detected, for the first time, more than 10 spectroscopic (and thus secure) close pairs of galaxies at  high redshift ($z>4$). The peak around $z=1$ for the UDF-Mosaic is partially due to the presence of an over-dense structure at this redshift in the HUDF (Popesso et al., 2009; Table 2). The gap around $z=2$ is due to the well-known redshift desert of spectroscopic surveys in the optical (see also Inami et al.\,2017). Examples of close pairs of galaxies in each redshift bins chosen for the fraction computation (see \ref{sec:zbins}) are shown in Figures \ref{im_pair1}, \ref{im_pairs2}, and \ref{im_pairs3}.


\section{Stellar mass estimates and close pair classification}
\label{sect:masses}

\begin{figure}[t]
        \includegraphics[width=\columnwidth]{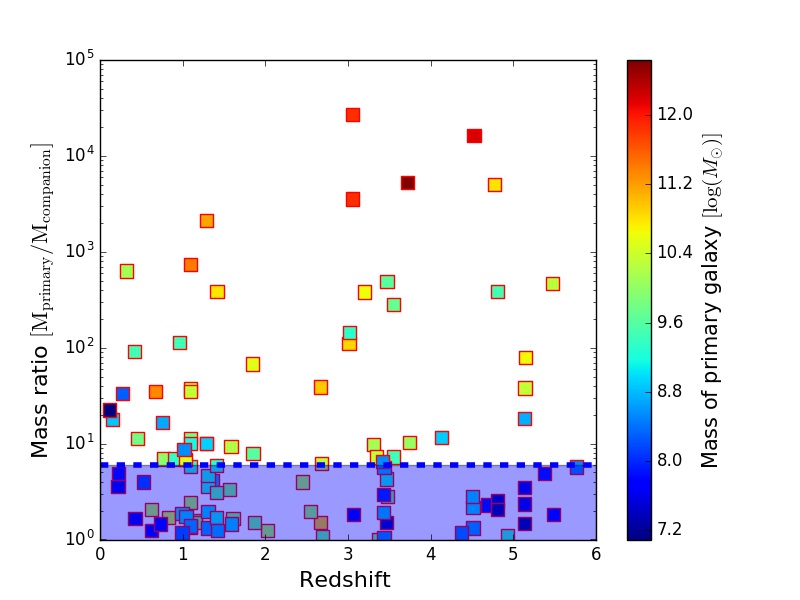}
   \caption{Stellar mass ratio of 113 close pairs identified in the MUSE deep fields as a function of redshift, colour coded with respect to the stellar mass of the primary galaxy. The blue dashed line indicates a mass ratio (primary over companion galaxy) limit of 6 chosen to distinguish major close pairs (blue coloured area) from minor close pairs.}
   \label{mass4}
 \end{figure}

The stellar mass ratio between galaxies in a close pair is a good proxy to distinguish between major and minor mergers, and hence to determine the associated fractions and rates. We thus used this proxy to isolate close pairs of galaxies with similar stellar masses and then fsed the subsequent analysis on this sample. We chose a mass ratio limit of 1:6 (defined as the ratio between the secondary and the primary galaxies) to really differentiate between the major and minor close pairs. This choice is justified by the fact that, with MUSE deep observations, we are probing a much broader range of galaxy masses than previous studies, allowing us to detect galaxy pairs with a mass ratio much lower than 1:4 at any redshift (see Fig.\,\ref{mass4}), which is the limit usually adopted in previous studies (see e.g.\, Lopez-Sanjuan et al.\,2013; Tasca et al.\,2014). 

We estimated the stellar masses of all the galaxies in the parent sample using the stellar population synthesis code FAST (Fitting and Assessment of Synthetic Templates; Kriek et al.\,2009); which fits model templates to the spectral energy distribution (SED) of galaxies based on the HST photometry, as described in Contini et al.\,(2016) for the HDF-S galaxies. For UDF-Mosaic and \textsf{udf-10}, we used the extended UV-to-NIR ACS and WFC3 photometry of Rafelski et al.\,(2015). We chose Bruzual \& Charlot (2003) for the stellar library, Calzetti et al.\,(2000) for the dust attenuation law, and a Chabrier (2003) initial mass function. 

Stellar masses of galaxies below $z\approx 3$ are well constrained with the UV-to-NIR photometry. However, stellar masses of higher redshift galaxies, derived with observed-frame UV-to-NIR photometry only, are known to be more uncertain. In order to increase the robustness of stellar mass estimates for high-redshift galaxies ($z\geqslant 3$) we used additional mid-infrared IRAC photometry from the {\it GOODS Re-ionization Era wide-Area Treasury from Spitzer} programme (GREATS; PI: Ivo Labbe), which provides the deepest data available over the MUSE-HUDF region. Photometry is measured using the software {\tt mophongo} (Labbe et al.\, 2015), which subtracts any neighbouring objects by a segmented, PSF-matched, HST image. This process is critical for accurate photometry because of the broader Spitzer IRAC PSF (see details in Lam et al, in prep). We further checked that the SED–derived mass ratios are consistent with the difference in near-infrared HST magnitudes of the two galaxies, as magnitudes in these bands can be considered as a rough proxy for stellar mass.

\begin{figure}[t]
        \includegraphics[width=\columnwidth]{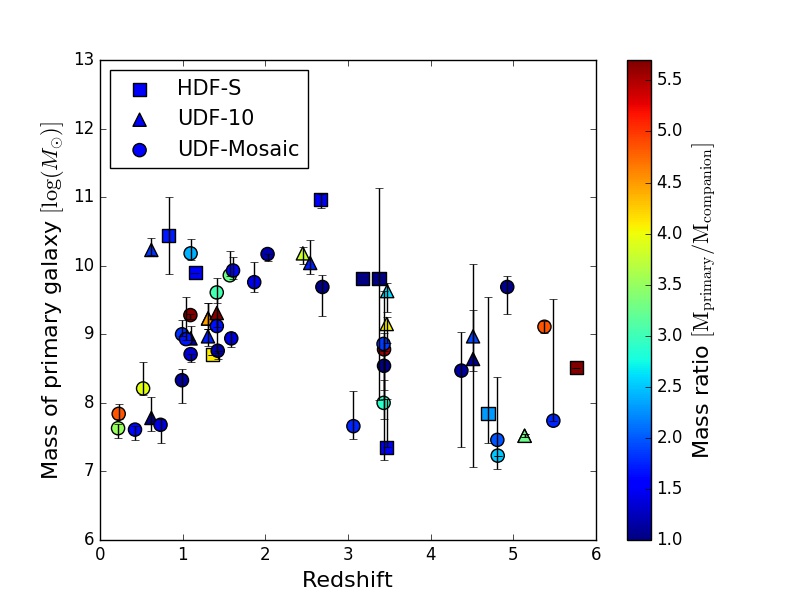}
   \caption{
  Stellar mass of the primary galaxy as a function of redshift for our major close pairs sample, colour coded with respect to the galaxy mass ratio in the pair. The primary galaxy is the more massive galaxy of the pair. The circles are pairs in the UDF-Mosaic, triangles in \textsf{udf-10}, and squares in HDF-S. Except in the redshift ``desert'' ($z\sim 1.5-2.8$), the mass range probed with MUSE observations does not change significantly with redshift, with a fairly good completeness level between $\approx 10^7-10^{10}$\Msun.}
   \label{fig:mass_vs_z}
 \end{figure}

With this sample of close pairs, as for the parent sample of galaxies, we probed a large domain of galaxy stellar masses in the range $\sim10^{7}-10^{11}$\Msun\ (see Fig.\,\ref{fig:mass_vs_z}), in which there is a high percentage of low-mass galaxies ($<10^{9.5}$\Msun), especially at very high redshift ($z>3$). From our sample of 113 close pairs (see Fig.\,\ref{mass4}), we identified a total of 54 major close pairs with a stellar mass ratio higher than 1:6. If we apply a mass ratio limit of 1:4, as in previous studies, we only lose eight pairs. But if we push this limit up to 1:10 we gain twenty-two pairs, as we are clearly entering into the minor merger regime. We checked that the relative number of identified close pairs scales roughly with the mass ratio, as expected from theory. To do so, we compared our measurements for two mass ratios regimes (major: $\leqslant 1/4$ and major$+$minor: $\leqslant 1/10$) to the most recent predictions from numerical simulations: Illustris (Rodriguez-Gomez et al.\, 2015) and EAGLE (Qu et al.\, 2017). The results are very consistent taking into account measurement uncertainties such as cosmic variance. We measured an increase of the fraction of close pairs by a factor 1.65 between the major (mass ratio $\leqslant 1/4$) and the major$+$minor ($\leqslant 1/10$) regime, which is in very good agreement with Illustris (factor of 1.5 to 2, see their Fig.\,7, top/middle panel) and EAGLE (factor of 1.5 to 1.8) predictions. The fact that the measured value from MUSE data is close to the lower limit predicted by the simulations may reflects an edge effect due to the sharp cut-off in the mass ratio threshold. But this effect is marginal and does significantly not affect the measured pair fractions.

The basic properties (such as redshift, stellar mass, projected separation, and velocity difference) for the sample of major galaxy close pairs identified in the three MUSE deep fields are given in Table\,\ref{table:1}.

\section{Redshift evolution of the galaxy major merger fraction}
\label{sec:fraction}

\subsection{Redshift bins}
\label{sec:zbins}

In order to estimate the evolution of the merger fraction and rate, we divided our redshift domain into five bins containing enough close pairs for statistical significance.  

The first redshift bin $0.2\leqslant z_{r}< 1$, corresponding to our lowest redshift range, contains 10 pairs of galaxies. The second bin, $1\leqslant z_{r}< 1.5$, extends up to the loss of the \oii\  emission-line doublet in the MUSE spectral range and contains 14 pairs. The third redshift bin $1.5\leqslant z_{r}< 2.8$ is associated with the well-known redshift desert, where we do not have bright emission line falling in the MUSE spectral range, except a few \ciii\ emitters (Maseda et al.\,2017). This bin includes 9 pairs. Above $z=2.8$, the vast majority of the galaxies are identified through their \lya\ emission. We divided this very high-redshift domain into two bins according to the distribution of close pairs, $2.8\leqslant z_{r}< 4$ and $4\leqslant z_{r}\leqslant 6$. These two last bins contain 10 and 13 pairs, respectively.

\subsection{Major merger fraction up to $z\approx 6$}
\label{sec:frac_evol}

  \begin{figure*}[t]
     \begin{tabular}{c}
     
       \includegraphics[width=1\textwidth]{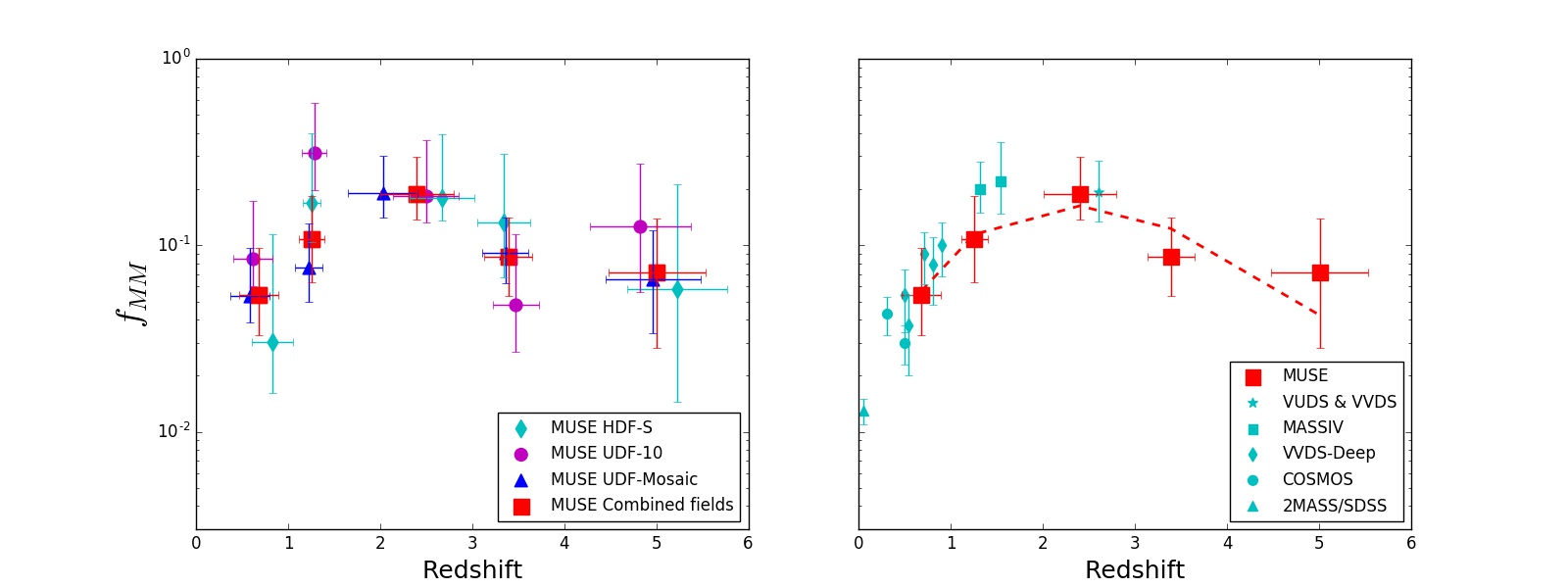}

        \end{tabular}
           \caption{Evolution of the galaxy major merger fraction up to $z\sim 6$. {\it Left}: Red squares correspond to the fraction for the combined analysis of the three MUSE fields. The other symbols indicate the estimates of the fraction from HDF-S, \textsf{udf-10} and UDF-Mosaic individually. {\it Right}: Combined major merger fractions from MUSE data (red squares) are compared to previous estimates (light blue symbols; de Ravel et al. 2009; Lopez-Sanjuan et al. 2011; Xu et al. 2012; Lopez-Sanjuan et al. 2013; Tasca et al. 2014). The dashed line is the least-squares fit of a combined power-law and exponential function, $f_{MM}\sim 0.056(1+z)^{5.910}e^{-1.814(1+z)}$, to the data.}
             \label{fig:merger_fraction}
   \end{figure*}

The merger fraction from a spectroscopic pair count is simply the number of pairs divided by the number of primary individual galaxies in the sample. However, as our observations are limited in volume and luminosity, we must correct the merger fraction from these selection effects and incompleteness (e.g.\,de Ravel et al.\,2009).

Similarly to the relation used, for example, in Lopez-Sanjuan et al.\, (2013), the major merger fraction for a chosen redshift bin $z_{r}$ is defined as
\begin{equation} 
f_{\mathrm{MM}}(z_{r})=\frac{N_p^{\mathrm{corr}}}{N_\mathrm{g}^{\mathrm{corr}}}=C_1\frac{\sum\limits_{K=1}^{N_p} \frac{\omega_{z}^{K_1} }{C_2(z_{r})}\frac{\omega_{z}^{K_2} }{C_2(z_{r})} \omega_{A}^K}{\sum_{i=1}^{N_g} \frac{\omega_{z}^i }{C_2(z_{r})} } ,\end{equation}
where $N_g$ is the number of primary galaxies in the parent sample; $N_p$ is the number of major close pairs; $C_1$ accounts for the missing companions due to our limit in spatial resolution (see sect.\,\ref{sec:limitations}); $\omega_{C}^K$ is the redshift confidence weight, which takes into account the confidence in the $z$ measurement (e.g.\,Inami et al.\, 2017); $\omega_{A}$ is the area weight, which takes into account that some galaxies are located on the border of the MUSE field of view; and finally $C_2(z_{r})$ is a correction term for the redshift incompleteness. \newline
All these terms are defined as
\begin{itemize}
\item [$\bullet$] $C_1=\frac{(r_p^{max})^{2} }{(r_p^{max})^{2}-(r_p^{min})^{2}}$ \newline

\item [$\bullet$] $\omega_{z}^K$, the redshift confidence weight, 
$$\omega_{z}=
\begin{cases}1 & \mbox{ if } z_{conf} =3 \text{ or } 2 \\
0.6 & \mbox{ if } z_{conf} =1. \\
\end{cases}$$
A maximum value of 1 is chosen for the weight of secure redshifts (with confidence of 3 or 2). To reduce the influence of unsecure pairs, i.e.\, with one of the galaxy flagged with a redshift confidence of 1, a weight of 0.6 is applied (i.e.\, we are 60\% sure of the redshift estimate). Varying this value in the range $0.5-0.7$ has almost no impact on the final fractions.

\item [$\bullet$] $\omega_{A}$, the area weight

$$\omega_{A} = \frac{A_{rp}}{A_{\rm MUSE}},$$ where $A_{rp}$ is the area of a circle of radius $r_p^{max}$ and $A_{\rm MUSE}$ is the corresponding area in the MUSE data cubes. This term has a very low impact on the fraction.

\item [$\bullet$] $C_2(z_{r})$ corrects for the spectroscopic redshift incompleteness and is defined, in each field and redshift bin, as the number of spectroscopic redshifts divided by the number of photometric redshifts, estimated in Brinchmann et al.\,(2017). We assumed that the photometric redshift measurements are uniformly representative of the true redshift distribution.
For galaxies at $z\leqslant 1.5$, and $1.5<z<2.8$, we applied a magnitude cut of $F775W\leqslant 29$ and $27$ mag on the parent sample, corresponding to the magnitude limit for the spectroscopic redshift identification of galaxies in these redshift intervals (see Inami et al. 2017). This concerns galaxies at $z\leqslant2.8$ only, since the emission-line source detection method using ORIGIN (see details in Bacon et al. 2017) identifies fainter objects for $z\geqslant 2.8$. Moreover, since the photometry in the \textsf{udf-10} has a much larger multi-wavelength coverage compared with the HDF-S, and these two fields have approximately the same sensitivity with MUSE (factor of 1.6 better for \textsf{udf-10}; Bacon et al.\,2017), we used the \textsf{udf-10} completeness corrections for the HDF-S. Values for these completeness corrections are listed in Table.\ref{table:2}. As expected, at high redshift, the completeness is higher in the \textsf{udf-10} than in the UDF-Mosaic, which is consistent with the difference in depth between these two fields. Up to $z\sim1.5$, we are almost 50\% complete both for the deep fields and the medium-deep UDF-mosaic. The completeness decreases between $z\approx 1.5$ and $z\approx 2.8$, corresponding to the redshift ``desert'' and stays almost constant over the two last redshift bins at approximately 40-50\% and 20-30\%, respectively, for the \textsf{udf-10} and UDF-Mosaic.

\end{itemize}
The error budget on the merger fraction was obtained by combining a purely statistical error on the estimated fractions  and an error due to the cosmic variance. We derived the statistical error as a confidence interval from a Bayesian approach (see e.g.\,Cameron 2011). The cosmic variance is a term inherent to observational studies and translates the impact of cosmic large-scale structures in measurements. We applied  the recipes of Moster et al.\,(2011) to compute the total cosmic variance (see Table\,\ref{table:2}) for the two uncorrelated fields: the HDF-S and UDF-Mosaic. This depends strongly on the geometry and volume of each field and on the redshift and mass bins assumed. For $z\leqslant 2$, it does not have a great influence since the uncertainties due to the cosmic variance are below 20\%. For this redshift range the error budget is  dominated by the low statistics. Whereas for $z\geqslant 3$, the cosmic variance predominates with uncertainties up to $\approx 50$\%.

We estimated the fraction of major close pairs for each field individually and for the combined study of the three MUSE fields put together (see Fig.\,\ref{fig:merger_fraction}, left and right panels, respectively). 
The comparison of the fractions for the individual fields clearly shows the effect of the cosmic variance. However, taking into account error bars, the measurements in the individual fields are in good agreement over the five redshift bins. As more than half of the pairs are detected in the UDF-Mosaic, this field has a higher weight on the combined fraction than the other two deeper but smaller fields. Table\,\ref{table:2} summarizes, for each redshift bins, the completeness correction factors, error due to cosmic variance, median values of stellar masses, and number and fraction of major close pairs.

In Figure\,\ref{fig:merger_fraction} (right panel), we compared our estimates with previous results from the literature, restricting the comparison to other samples of close pairs robustly identified with spectroscopic studies. Similar values for separation limits, i.e. $r_p^{\rm max}=20-30\mathrm{h^{-1}kpc}$ and $\Delta v_{max}\sim 500$ \kms, were used in the MASSIV (Lopez-Sanjuan et al.\, 2013), VVDS/VUDS  (Tasca et al.\, 2014), and VVDS-deep (de Ravel et al.\, 2009; Lopez-Sanjuan et al.\, 2011) analyses to select close pairs. A typical mass ratio limit of 1:4 for major-merger pairs is usually adopted, except in de Ravel et al.\, (2009) who choose a magnitude difference limit of 1.5 mag between pair members. The major close pairs selection in the 2MASS/SDSS and COSMOS samples (Xu et al.\, 2012) follow approximately the same criteria with $5 \leqslant r_\mathrm{p} \leqslant 20 \mathrm{h^{-1}kpc}$ but with a lower mass ratio limit of 1:2.5. We must however keep in mind that the comparison is not so straightforward as the close pairs detected in the MUSE fields involve galaxies spread over a large range of stellar masses ($\sim 10^7-10^{11}$\Msun; see sect.\,\ref{sect:masses}), whereas the close pairs analysed so far in the literature involve massive galaxies only ($\geqslant 10^{10}$\Msun).

However, the major merger fractions estimated in the MUSE fields are in good agreement with those derived from previous analyses in similar redshifts, with a constant increase of the merger fraction with look-back time up to $z\approx 3$. At higher redshift, the fraction seems to decrease slowly or flatten down. 

 Since we chose a mass ratio limit of 1:6 to define our major close pair sample, some pairs could be missed at $z\geq 3$  owing to the non-detection of the companion of a primary galaxy with a very low stellar mass, i.e.\,with $\mathrm{M}_* \approx 10^{7}-10^{8}$\Msun.
  Consequently, we might probe a different mass regime at low and high redshifts.
However, as shown in Fig.\,\ref{mass4}, 
we detect close pairs at $z\geq 3$ with a mass ratio $\leq$1:4 and a primary galaxy stellar mass around $\mathrm{M}_* \approx 10^{7}-10^{8}$\Msun, as in the lower redshift range ($z\leq 1.5$). It is also clear from Fig.\,\ref{mass4} that for a mass ratio lower than 1:6; i.e.\,in the minor close pair regime, the primary galaxy stellar mass range for $z\leqslant 3$ galaxies is comparable to that for $z\geqslant 3$. 
We further checked that the evolutionary trend seen in Fig.\,\ref{fig:merger_fraction} does not change if the mass ratio threshold used to define our major close pair sample is set to a value of 1:3 or 1:4. Such a trend has a low impact on the estimate of the fraction, with a decrease of the fraction of $\approx 3$\% on average between a mass ratio limit of 1:6 and 1:3, but the evolution remains consistent. The conclusions are the same if we increase the lower limit of the primary galaxy stellar mass to $10^8$\Msun. 


\begin{table*}[h]
\caption{Major merger fractions up to $z\approx 6$ from the HDF-S, \textsf{udf-10} and UDF-Mosaic combined analysis. Cols.\,(1) and (2): Range of the redshift bin and its associated mean redshift for the close pairs sample. Cols.\,(3) and (4): Weight corresponding to spectroscopic redshift completeness for the two deep fields, based on the \textsf{udf-10}, $C_1(z_{r})$, and the UDF-Mosaic, $C_2(z_{r})$. Col.\,(5): Total cosmic variance for the combined field study, depending on the redshift bin and the median of stellar masses for the close pairs. Cols.\,(6) and (7): Median values of stellar masses for the parent and pairs samples respectively. Cols.\,(8) and (9): Number of galaxies, $N_g$, and pairs, $N_p$, for the redshift bin.  Col.\,(10): Major merger fraction.}
\label{table:2}      
\centering                          
\begin{tabular}{c c c c c c c c c c c c}        
\hline\hline  
&&&&&&&\\
 $z_r$ & $\overline{z_r}$&$C_1(z_{r})$&$C_2(z_{r})$&$\sigma_{v}$&$ \overline{M^{\star}_g}$ & $ \overline{M^{\star}_p}$ & $N_g$ & $N_p$& $f_{MM}$ \\
-&-&-&-&-&[log(\Msun)]&[log(\Msun)]&-&-&-\\
(1) & (2) & (3) & (4) & (5) & (6) & (7)&(8)&(9)&(10) \\

\hline                        
&&&&&&&&\\
$0.2\leqslant z< 1$&0.68 &0.48&0.45&0.15&8.21& 8.03& 404 &10 &$0.054_{-0.021}^{+0.042}$\\
$1\leqslant z< 1.5$&1.25 &0.45&0.44&0.18&8.96& 9.17& 297 &14 &$0.107_{-0.044}^{+0.076}$\\
$1.5\leqslant z< 2.8$&2.35&0.43&0.30 &0.15&9.58& 9.93& 152 & 9 &$0.188_{-0.051}^{+0.110}$\\
$2.8\leqslant z< 4$&3.39 &0.42&0.20&0.36&8.58& 8.82& 399& 10&$0.087_{-0.033}^{+0.054}$\\
$4\leqslant z\leqslant 6$&4.99&0.55&0.35&0.52&8.36& 7.91& 382 &  13& $0.072_{-0.043}^{+0.068}$\\
&&&&&&\\
\hline  
\end{tabular}
\end{table*}

\subsection{Separation by stellar mass}
\label{sect:mass_cut}

  \begin{figure*}[h]
    \begin{tabular}{c}       \includegraphics[width=1\textwidth]{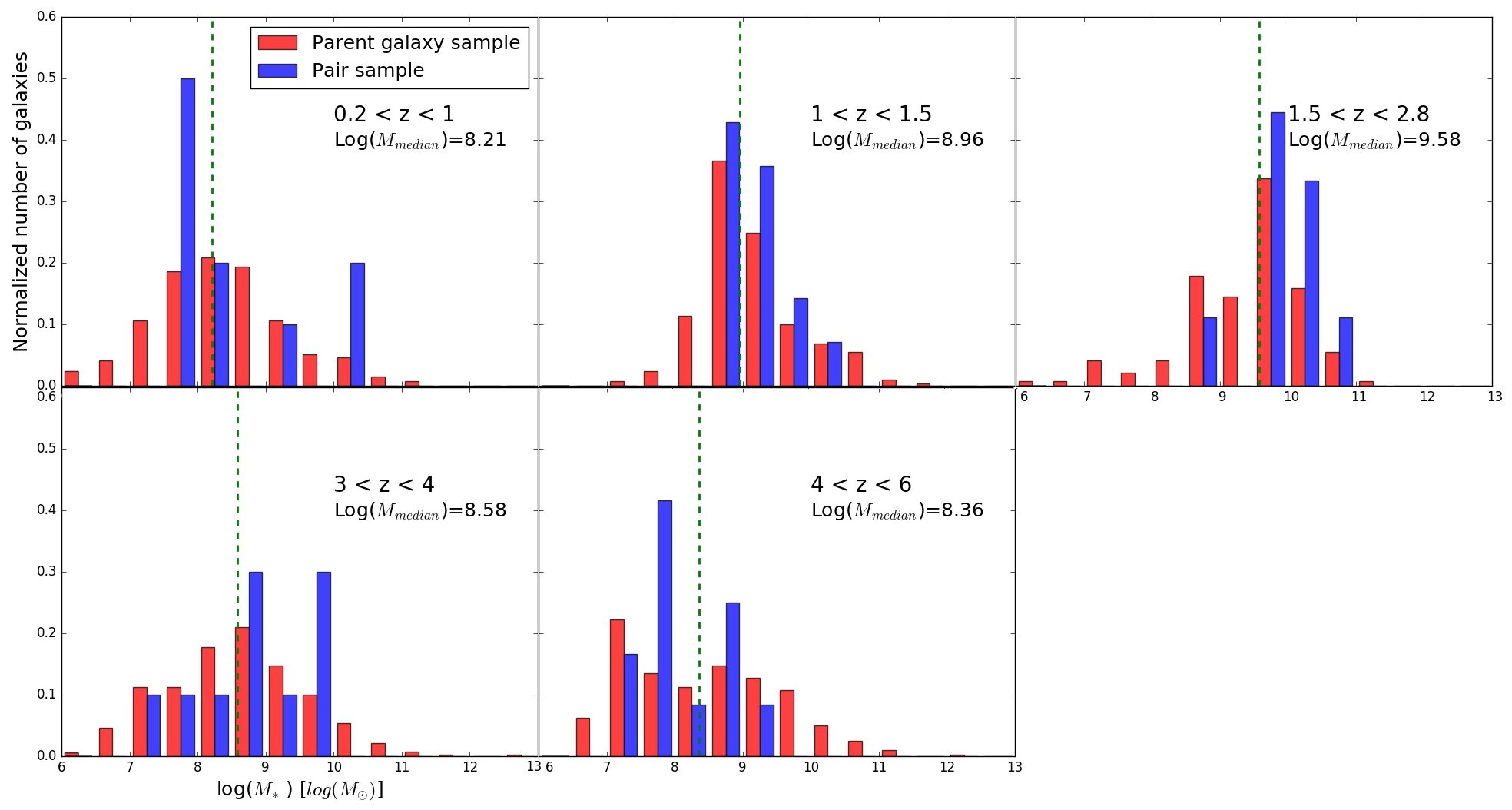}

        \end{tabular}
           \caption{ Stellar mass distribution of the parent (red) and close pair (blue) samples in each redshift bins. The reported median value of the parent sample is represented by the dashed green line. The distributions are normalized to the sum of stellar mass bins.}
             \label{fig:zbins}
   \end{figure*}
   
Figure\,\ref{fig:zbins} shows the normalized stellar mass distributions of the parent and close pair samples in each redshift bins. At all redshifts and stellar masses of the parent sample extend over four orders of magnitude from $\sim 10^7$\Msun\ to $\sim 10^{11}$\Msun. With median values between $10^8$\Msun\ to $10^9$\Msun\ (see Table\,\ref{table:2}), it is clear that with MUSE we are probing a lower mass domain than previous spectroscopic surveys, which pre-selected the targets according to their apparent magnitude. The only exception is the bin corresponding to the redshift desert, with a median mass above $10^9$\Msun, in agreement with the fact that most of the galaxies identified in this redshift range have a bright continuum.
The stellar mass distributions of galaxies in close pairs broadly follow the distributions of the parent sample. However, we have not found major close pairs made of very low-mass galaxies  (i.e.\,$\leqslant 10^{7.5}$\Msun) below $z\sim 3$, nor pairs of massive galaxies (i.e.\,$\geqslant 10^{10}$\Msun) above this redshift.

\begin{figure*}[t]
    \begin{tabular}{c}
       \includegraphics[width=1\textwidth]{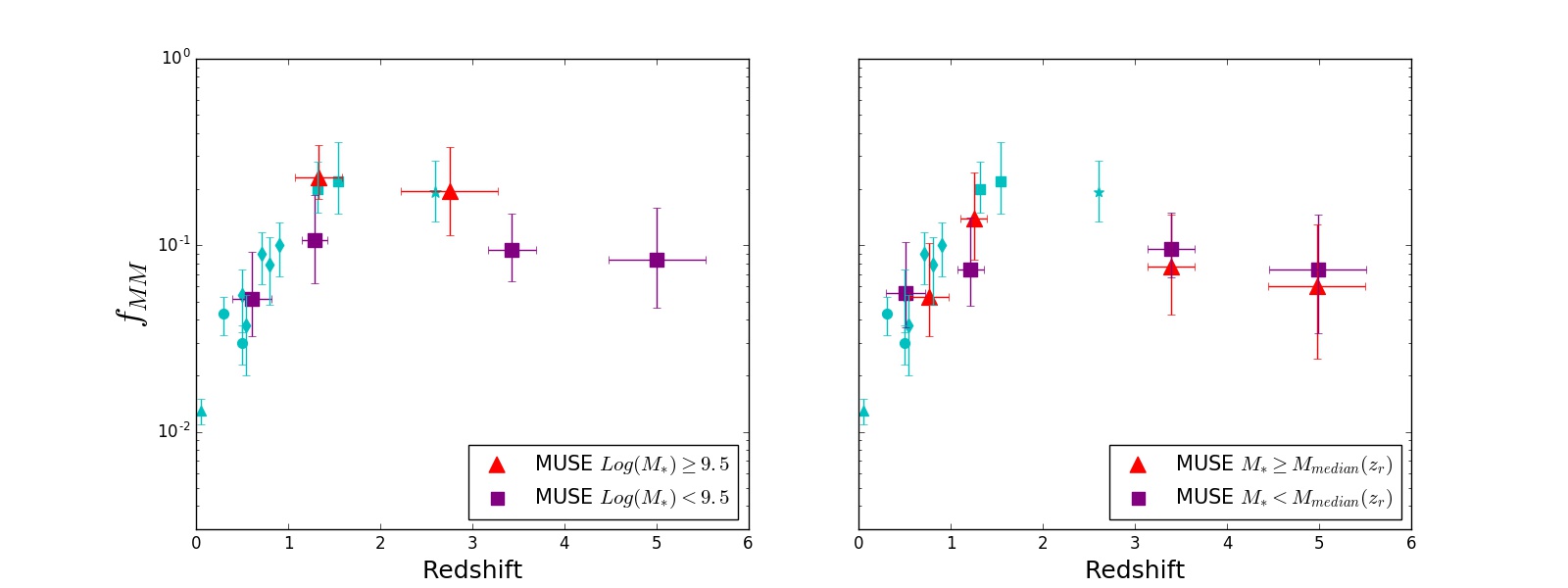}
        \end{tabular}
           \caption{Evolution of the major merger fraction for two ranges of stellar mass, assuming a constant separation limit of \Mstar\ $= 10^{9.5}$\Msun\ ({\it left panel}) or adopting the median value of stellar mass in each redshift bin as the separation limit ({\it right panel}). The purple squares and red triangles show the MUSE estimates for low-mass and massive galaxies, respectively. Previous estimates from the literature are shown with light blue symbols (see Fig.\,\ref{fig:merger_fraction} for references).}
             \label{fig:merger_fraction_3}
   \end{figure*}

An attempt to separate our sample of close pairs in stellar masses is shown in Fig.\,\ref{fig:merger_fraction_3}. We use the stellar mass of the primary galaxy to discriminate the pairs and test two different stellar mass limit criteria.

First, a constant stellar mass limit of $10^{9.5}$\Msun\ is chosen to distinguish low mass from massive galaxies over the entire redshift range (Fig.\,\ref{fig:merger_fraction_3}, left panel). 
For this analysis, the redshift bins defined previously (see sect.\,\ref{sec:zbins}) are modified to keep a significant statistic. We thus remove the bin corresponding to the redshift desert for the low-mass sample, and we define three new redshift bins $0.2\leqslant z_{r1}< 1$, $1\leqslant z_{r2}< 2$ and $2\leqslant z_{r3}\leqslant 4$ for the sample of ``massive'' galaxies (see Table\,\ref{table:3}).  As we have two pairs only in the first redshift bin, this data point is not shown in Fig.\,\ref{fig:merger_fraction_3} (left panel) but is still reported in Table\,\ref{table:2}.

The major merger fractions estimated for the high-mass samples are, within uncertainties, fairly consistent with previous works, with an increase of the fraction up to 23\% and 19\% at $z\approx 1.3$ and 2.7. The major merger fraction evolution of the low-mass sample seems to follow the same trend with a monotonically increases up to $z\sim 1.3-3,$ where it reaches a maximum of 11\% and then flattens or slightly decreases to 8-9\% between $3\leqslant z\leqslant 6$ (see Table\,\ref{table:3}). 

Since we probe a particularly low-mass regime in stellar masses with MUSE, a second approach is to define the mass limit as the median value of the mass distribution for the parent galaxy sample. This limit varies with redshift, as described in section\,\ref{sec:zbins}. With this separation, the two close pairs samples are more evenly distributed. Figure\,\ref{fig:merger_fraction_3} (right panel) shows a trend similar to the left panel with small differences between the two estimates of the major merger fraction according to these median mass limits. Overall, the major close pair fraction for low-mass and massive galaxies follow the same trend.
However, there is a potential reverse trend between the two mass bins in this figure, even if the uncertainties on the fraction do not allow any firm conclusion. Indeed, around $z\approx 1.5$, the merger fraction is higher for massive galaxies than for low-mass galaxies, but at higher redshift ($z\geqslant 3$) this trend is reversed, as seen in some simulations (e.g.\, Qu et al.\,2017).

\subsection{Comparison with recent simulations}
\label{sec:comp_simus}

We can compare our merger fractions to predictions from hydrodynamic simulations that model the dark matter and baryonic components of a cosmological volume consistently. Until recently, there have been very few attempts (e.g. Maller et al. 2006) to determine the evolution of galaxy merger fractions using such simulations because it was not possible to produce large enough samples of realistic galaxies. This situation greatly improved over the last years with simulations such as HORIZON-AGN (Dubois et al. 2014), Illustris (Vogelsberger et al.\,2014), and EAGLE (Crain et al.\,2015; Schaye et al.\,2015). 

A straightforward comparison with observations is to measure the close pair fraction directly from the simulations, to be compared to observations without having to make any assumption about the merger timescales (see sect.\,\ref{sec:intro}). 
Estimates of the major merger fraction evolution with redshift are available from the HORIZON-AGN (Kaviraj et al.\,2015), EAGLE (Qu et al.\,2017), and Illustris (Snyder et al.\,2017) simulations. 

Using the EAGLE simulations, Qu et al. (2017) have built merger trees to connect galaxies to their progenitors. From snapshots at different redshifts, they searched for pairs of galaxies following selection criteria similar to those used in observational close-pair analysis, such as the separation distance and mass ratio of the galaxies. Estimates of the major close pairs  fraction are given in three stellar mass ranges up to redshift $\approx 4$.
This fraction increases monotonically before leveling off at $z=1.5-3$ and even declines for the most massive galaxies. This trend is best fitted with a combined power-law and exponential function. 

Based on the HORIZON-AGN simulation, Kaviraj et al.\,(2015) have probed the merger histories of massive galaxies and predicted the fractions of galaxy pairs in the redshift range $1<z<4$ and various mass ratios. The trend is roughly similar to predictions by EAGLE in the same redshift range with a flat increase of the merger fraction up to $z\approx 3$ and then a decrease towards higher redshift.

From the Illustris simulation, Snyder et al.\,(2017) have created three synthetic light cone catalogues and measured  pair fractions using a velocity criterion inspired by photometric redshift precision in deep surveys, i.e.\, $\Delta v_{\rm max} = 18000$ \kms\ at $z=2$. The fraction seems to be roughly flat between $z\approx 0.5-3$ and then decreases up to $z\approx 4$. However this trend requires a decreasing observability timescale with redshift, which corresponds to the timescale at which a close pair can be identified in a snapshot catalogue. 

\begin{figure}[t]
        \includegraphics[width=\columnwidth]{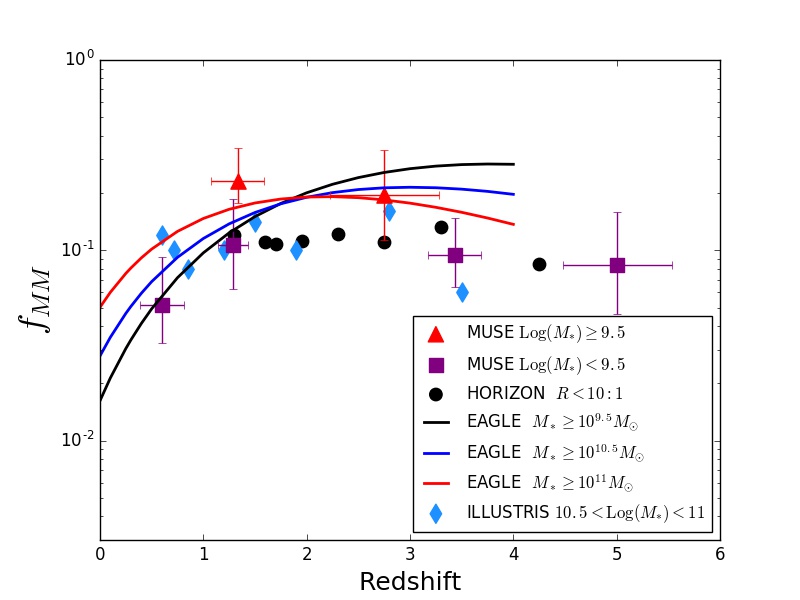}
   \caption{Major merger fraction compared to recent numerical simulations. Symbols with error bars are estimates from our MUSE sample divided into low-mass ($\leqslant 10^{9.5}$\Msun; purple squares) and massive ($> 10^{9.5}$\Msun; red triangles) galaxies. The black points indicate the predictions from the HORIZON-AGN simulation (Kaviraj et al.\,2015) and correspond to the pair fraction for massive galaxies ($\geqslant 10^{10}$\Msun) with a mass ratio between the primary and companion galaxy that is lower than 10:1. The solid lines indicate estimates from the EAGLE simulations for three galaxy stellar mass ranges. For these predictions a combined power-law and exponential fitting function, $f_{MM}\sim a(1+z)^{b}e^{-c(1+z)}$, was used (see Qu et al.\,2017 for details). Finally, the blue diamonds correspond to the major pair fraction for massive galaxies in the ILLUSTRIS simulation (Snyder et al.\, 2017).}
   \label{rate_4}
 \end{figure}

Figure\,\ref{rate_4} compares the predictions from these simulations to our major merger fraction estimates. Even though the simulated samples are biased towards more massive galaxies than studied in this work, the trend of the fraction evolution in these simulations is consistent with our study, especially when pairs of both low- and high-mass galaxies, which have stellar mass ratios down to $\sim$ 1:10, are taken into account in the simulations.

 \begin{table*}
\caption{Major merger fraction  up to $z\approx 6$ from MUSE observations for different redshift and stellar mass intervals. Cols.\,(1) and (2): Range of the redshift bin and its associated mean redshift for the close pairs sample. Cols\,(3): Median value of stellar mass of the pairs sample. Cols.\,(4) and (5): Number of pairs, $N_p$, and galaxies, $N_g$.  Cols\,(6): Major merger fraction.}
\label{table:3}      
\centering                          
\begin{tabular}{c c c c c c c }        
\hline\hline  
&&&&&&\\
 $z_r$ & $\overline{z_r}$& $ \overline{M^{\star}_p}$ & $N_p$ & $N_g$& $f_{MM}$ \\
-&-&[log(\Msun)]&-&-&-\\
(1) & (2) & (3) & (4) & (5) & (6)  \\

\hline                        
&&&&&&\\
 &&&&$M^{\star}<M_{\rm median}(z_r)$&\\
 &&&&&&\\
$0.2\leqslant z< 1$&0.51& 7.68& 5 &207 &$0.055_{-0.019}^{+0.048}$\\
$1\leqslant z\leqslant 1.5$&1.21 & 8.76& 5 &153 &$0.074_{-0.027}^{+0.066}$\\
$3\leqslant z< 4$&3.39 & 7.86& 4 &211 &$0.096_{-0.029}^{+0.054}$\\
$4\leqslant z\leqslant 6$&4.98 & 7.52& 7 &223 &$0.074_{-0.040}^{+0.071}$ \\
&&&&&&\\
\hline  
&&&&&&\\
 &&&&$M^{\star}\geqslant M_{\rm median}(z_r)$&\\
 &&&&&&\\
$0.2\leqslant z< 1$&0.76& 9.00& 4 &197 &$0.053_{-0.020}^{+0.050}$ \\
$1\leqslant z\leqslant 1.5$&1.24 &9.28& 9 &146 &$0.139_{-0.055}^{+0.0107}$ \\
$3\leqslant z< 4$&3.39 & 9.45& 6 &188 &$0.077_{-0.034}^{+0.067}$ \\
$4\leqslant z\leqslant 6$&4.86 & 8.83& 6 &197 &$0.060_{-0.035}^{+0.068}$ \\
&&&&\\
\hline  
&&&&\\
 &&&& log($M^{\star}$)$< 9.5$&\\
 &&&&&&\\
$0.2\leqslant z< 1$&0.60&7.81& 8 &357 &$0.052_{-0.019}^{+0.040}$ \\
$1\leqslant z\leqslant 1.5$&1.28 &8.97& 11 &230 &$0.106_{-0.044}^{+0.081}$ \\
$3\leqslant z< 4$&3.43 &8.54& 7 &329 &$0.095_{-0.031}^{+0.052}$ \\
$4\leqslant z\leqslant 6$&4.99 & 7.91& 13 &344 &$0.083_{-0.037}^{+0.076}$ \\
 &&&&&&\\
 \hline
&&&&&&\\
 &&&& log($M^{\star}$)$\geqslant 9.5$&\\
 &&&&&&\\
 $0.2\leqslant z< 1$&0.73& 10.34& 2 &47 &$0.071_{-0.018}^{+0.105}$ \\
$1\leqslant z< 2$&1.33 &9.88& 6 &112 &$0.232_{-0.056}^{+0.112}$\\
$2\leqslant z\leqslant 4$&2.75 & 9.94& 8 &118 &$0.195_{-0.081}^{+0.142}$ \\
 &&&&&&\\
 \hline
\end{tabular}
\end{table*}

\section{Summary and conclusions}
\label{sec:conclusion}

We used deep MUSE observations in the HUDF and HDF-S to identify 113 secure close pairs of galaxies among a parent sample of 1801 galaxies spread over a large redshift range ($0.2<z<6$) and stellar masses ($10^7-10^{11}$\Msun), thus probing about 12 Gyr of galaxy evolution. We used stellar masses derived from SED fitting to isolate a sample of 54 major close pairs with a galaxy mass ratio limit of 1:6. Thanks to this exquisite data set, we provided, for the first time, robust observational constraints on the galaxy major merger fraction up to $z\approx 6$ using spectroscopic close pair counts. 

Among this sample of major close pairs, we identified 20 systems at  high redshift ($z\geqslant 3$) through their \lya\ emission. For these galaxies, we used the FWHM of the \lya\ emission line as a proxy to retrieve their systemic redshift, following theoretical and observational arguments recently developed in Verhamme et al.\,(2017). The sample of major close pairs was divided into five redshift intervals to probe the evolution of the merger fraction with cosmic time. Our estimates are in very good agreement with previous close pair counts with a constant increase of the merger fraction up to $z\approx 3,$ where it reaches a maximum of 20\%. At higher redshift, we show  that the fraction slowly decreases down to about 10\% at $z\approx6$. 

We further divided the sample into two ranges of stellar masses using either a constant separation limit of $10^{9.5}$\Msun\ or the median value of stellar mass computed in each redshift bin. We show that there is a potential reversed trend between the cosmic evolution of the merger fraction in these two mass regimes. Indeed, around $z\approx 1.5$, the merger fraction is higher for massive galaxies, but at higher redshift ($z\geqslant 3$) this trend is reversed. The cosmic evolution of these new estimates of the major merger fraction up to $z\approx6$ is in agreement with recent predictions of cosmological numerical simulations, such as HORIZON-AGN (Kaviraj et al.\,2015), EAGLE (Qu et al.\,2017), and Illustris (Snyder et al.\,2017). 


The shape of the cosmic evolution of the galaxy major merger fraction up to $z\approx 6$ derived from our MUSE data set, which shows an increase up to $z\approx3$ and then a decrease at higher redshifts, is reminiscent of the well-known cosmic star formation rate evolution (e.g.\, Madau \& Dickinson 2014). This similarity will be further investigated in subsequent papers, making use of larger MUSE data sets acquired over the course the Guaranteed Time Observations to better assess the role of mergers in the growth of galaxies over more than 12 Gyrs.

\begin{acknowledgements}
This work has been carried out thanks to the support of the ANR FOGHAR (ANR-13-BS05-0010-02), the OCEVU Labex (ANR-11-LABX-0060), and the A*MIDEX project (ANR-11-IDEX-0001-02) funded by the ``Investissements d'avenir'' French government programme. BE acknowledges financial support from “Programme National de Cosmologie et Galaxies” (PNCG) of CNRS/INSU, France. LW acknowledges funding by the Competitive Fund of the Leibniz Association through grant SAW-2015-AIP-2. PMW received support through BMBF Verbundforschung (project MUSE-AO, grant 05A14BAC). JB acknowledges support by Funda{\c c}{\~a}o para a Ci{\^e}ncia e a Tecnologia (FCT) through national funds (UID/FIS/04434/2013) and Investigador FCT contract IF/01654/2014/CP1215/CT0003, and by FEDER through COMPETE2020 (POCI-01-0145-FEDER-007672).TG is grateful to the LABEX Lyon Institute of Origins (ANR-10-LABX-0066) of the Université de Lyon for its financial support within the programme 'Investissements d'Avenir' (ANR-11-IDEX-0007) of the French government operated by the National Research Agency (ANR).
\end{acknowledgements}

\nocite{*}
\bibliographystyle{aa}
\bibliography{Biblio}

\begin{thebibliography}{98}
\expandafter\ifx\csname natexlab\endcsname\relax\def\natexlab#1{#1}\fi

\bibitem[{{Abraham} {et~al.}(1996){Abraham}, {van den Bergh}, {Glazebrook},
  {Ellis}, {Santiago}, {Surma}, \& {Griffiths}}]{ABG1996}
{Abraham}, R.~G., {van den Bergh}, S., {Glazebrook}, K., {et~al.} 1996, \apjs,
  107, 1

\bibitem[{{Alonso} {et~al.}(2004){Alonso}, {Tissera}, {Coldwell}, \&
  {Lambas}}]{ATC2004}
{Alonso}, M.~S., {Tissera}, P.~B., {Coldwell}, G., \& {Lambas}, D.~G. 2004,
  \mnras, 352, 1081

\bibitem[{{Bacon} {et~al.}(2015){Bacon}, {Brinchmann}, {Richard}, {Contini},
  {Drake}, {Franx}, {Tacchella}, {Vernet}, {Wisotzki}, {Blaizot}, {Bouch{\'e}},
  {Bouwens}, {Cantalupo}, {Carollo}, {Carton}, {Caruana}, {Cl{\'e}ment},
  {Dreizler}, {Epinat}, {Guiderdoni}, {Herenz}, {Husser}, {Kamann}, {Kerutt},
  {Kollatschny}, {Krajnovic}, {Lilly}, {Martinsson}, {Michel-Dansac},
  {Patricio}, {Schaye}, {Shirazi}, {Soto}, {Soucail}, {Steinmetz}, {Urrutia},
  {Weilbacher}, \& {de Zeeuw}}]{BBR2015}
{Bacon}, R., {Brinchmann}, J., {Richard}, J., {et~al.} 2015, \aap, 575, A75

\bibitem[{{Bacon} \& {et al.}(2017)}]{B2017}
{Bacon}, R. \& {et al.} 2017, \aap, submitted

\bibitem[{{Baugh}(2006)}]{B2006}
{Baugh}, C.~M. 2006, Reports on Progress in Physics, 69, 3101

\bibitem[{{Beckwith} {et~al.}(2006){Beckwith}, {Stiavelli}, {Koekemoer},
  {Caldwell}, {Ferguson}, {Hook}, {Lucas}, {Bergeron}, {Corbin}, {Jogee},
  {Panagia}, {Robberto}, {Royle}, {Somerville}, \& {Sosey}}]{BSK2006}
{Beckwith}, S.~V.~W., {Stiavelli}, M., {Koekemoer}, A.~M., {et~al.} 2006, \aj,
  132, 1729

\bibitem[{{Bell} {et~al.}(2008){Bell}, {Zucker}, {Belokurov}, {Sharma},
  {Johnston}, {Bullock}, {Hogg}, {Jahnke}, {de Jong}, {Beers}, {Evans},
  {Grebel}, {Ivezi{\'c}}, {Koposov}, {Rix}, {Schneider}, {Steinmetz}, \&
  {Zolotov}}]{BZB2008}
{Bell}, E.~F., {Zucker}, D.~B., {Belokurov}, V., {et~al.} 2008, \apj, 680, 295

\bibitem[{{Besla} {et~al.}(2012){Besla}, {Kallivayalil}, {Hernquist}, {van der
  Marel}, {Cox}, \& {Kere{\v s}}}]{BKH2012}
{Besla}, G., {Kallivayalil}, N., {Hernquist}, L., {et~al.} 2012, \mnras, 421,
  2109

\bibitem[{{Bluck} {et~al.}(2009){Bluck}, {Conselice}, {Bouwens}, {Daddi},
  {Dickinson}, {Papovich}, \& {Yan}}]{BCB2009}
{Bluck}, A.~F.~L., {Conselice}, C.~J., {Bouwens}, R.~J., {et~al.} 2009, \mnras,
  394, L51

\bibitem[{{Bluck} {et~al.}(2012){Bluck}, {Conselice}, {Buitrago},
  {Gr{\"u}tzbauch}, {Hoyos}, {Mortlock}, \& {Bauer}}]{BCB2012}
{Bluck}, A.~F.~L., {Conselice}, C.~J., {Buitrago}, F., {et~al.} 2012, \apj,
  747, 34

\bibitem[{{Brinchmann} {et~al.}(1998){Brinchmann}, {Abraham}, {Schade},
  {Tresse}, {Ellis}, {Lilly}, {Le F{\`e}vre}, {Glazebrook}, {Hammer},
  {Colless}, {Crampton}, \& {Broadhurst}}]{BAS1998}
{Brinchmann}, J., {Abraham}, R., {Schade}, D., {et~al.} 1998, \apj, 499, 112

\bibitem[{{Bruzual} \& {Charlot}(2003)}]{BC2003}
{Bruzual}, G. \& {Charlot}, S. 2003, \mnras, 344, 1000

\bibitem[{{Bundy} {et~al.}(2005){Bundy}, {Ellis}, \& {Conselice}}]{BEC2005}
{Bundy}, K., {Ellis}, R.~S., \& {Conselice}, C.~J. 2005, \apj, 625, 621

\bibitem[{{Bundy} {et~al.}(2009){Bundy}, {Fukugita}, {Ellis}, {Targett},
  {Belli}, \& {Kodama}}]{BFE2009}
{Bundy}, K., {Fukugita}, M., {Ellis}, R.~S., {et~al.} 2009, \apj, 697, 1369

\bibitem[{{Calzetti} {et~al.}(2000){Calzetti}, {Armus}, {Bohlin}, {Kinney},
  {Koornneef}, \& {Storchi-Bergmann}}]{CAB2000}
{Calzetti}, D., {Armus}, L., {Bohlin}, R.~C., {et~al.} 2000, \apj, 533, 682

\bibitem[{{Cameron}(2011)}]{C2011}
{Cameron}, E. 2011, \pasa, 28, 128

\bibitem[{{Casteels} {et~al.}(2014){Casteels}, {Conselice}, {Bamford},
  {Salvador-Sol{\'e}}, {Norberg}, {Agius}, {Baldry}, {Brough}, {Brown},
  {Drinkwater}, {Driver}, {Graham}, {Bland-Hawthorn}, {Hopkins}, {Kelvin},
  {L{\'o}pez-S{\'a}nchez}, {Loveday}, {Robotham}, \&
  {V{\'a}zquez-Mata}}]{CCB2014}
{Casteels}, K.~R.~V., {Conselice}, C.~J., {Bamford}, S.~P., {et~al.} 2014,
  \mnras, 445, 1157

\bibitem[{{Chabrier}(2003)}]{C2003}
{Chabrier}, G. 2003, \pasp, 115, 763

\bibitem[{{Conselice}(2014)}]{C2014}
{Conselice}, C.~J. 2014, \araa, 52

\bibitem[{{Conselice} {et~al.}(2003){Conselice}, {Bershady}, {Dickinson}, \&
  {Papovich}}]{CBD2003}
{Conselice}, C.~J., {Bershady}, M.~A., {Dickinson}, M., \& {Papovich}, C. 2003,
  \aj, 126, 1183

\bibitem[{{Conselice} {et~al.}(2000){Conselice}, {Bershady}, \&
  {Jangren}}]{CBJ2000}
{Conselice}, C.~J., {Bershady}, M.~A., \& {Jangren}, A. 2000, \apj, 529, 886

\bibitem[{{Conselice} {et~al.}(2014){Conselice}, {Bluck}, {Mortlock},
  {Palamara}, \& {Benson}}]{CBM2014}
{Conselice}, C.~J., {Bluck}, A.~F.~L., {Mortlock}, A., {Palamara}, D., \&
  {Benson}, A.~J. 2014, \mnras, 444, 1125

\bibitem[{{Conselice} {et~al.}(2011){Conselice}, {Bluck}, {Ravindranath},
  {Mortlock}, {Koekemoer}, {Buitrago}, {Gr{\"u}tzbauch}, \& {Penny}}]{2CBR2011}
{Conselice}, C.~J., {Bluck}, A.~F.~L., {Ravindranath}, S., {et~al.} 2011,
  \mnras, 417, 2770

\bibitem[{{Conselice} {et~al.}(2008){Conselice}, {Rajgor}, \&
  {Myers}}]{CRM2008}
{Conselice}, C.~J., {Rajgor}, S., \& {Myers}, R. 2008, \mnras, 386, 909

\bibitem[{{Conselice} {et~al.}(2009){Conselice}, {Yang}, \& {Bluck}}]{CYB2009}
{Conselice}, C.~J., {Yang}, C., \& {Bluck}, A.~F.~L. 2009, \mnras, 394, 1956

\bibitem[{{Contini} {et~al.}(2016){Contini}, {Epinat}, {Bouch{\'e}},
  {Brinchmann}, {Boogaard}, {Ventou}, {Bacon}, {Richard}, {Weilbacher},
  {Wisotzki}, {Krajnovi{\'c}}, {Vielfaure}, {Emsellem}, {Finley}, {Inami},
  {Schaye}, {Swinbank}, {Gu{\'e}rou}, {Martinsson}, {Michel-Dansac},
  {Schroetter}, {Shirazi}, \& {Soucail}}]{CEB2016}
{Contini}, T., {Epinat}, B., {Bouch{\'e}}, N., {et~al.} 2016, \aap, 591, A49

\bibitem[{{Cooke} {et~al.}(2010){Cooke}, {Berrier}, {Barton}, {Bullock}, \&
  {Wolfe}}]{CBB2010}
{Cooke}, J., {Berrier}, J.~C., {Barton}, E.~J., {Bullock}, J.~S., \& {Wolfe},
  A.~M. 2010, \mnras, 403, 1020

\bibitem[{{Crain} {et~al.}(2015){Crain}, {Schaye}, {Bower}, {Furlong},
  {Schaller}, {Theuns}, {Dalla Vecchia}, {Frenk}, {McCarthy}, {Helly},
  {Jenkins}, {Rosas-Guevara}, {White}, \& {Trayford}}]{2CSB2015}
{Crain}, R.~A., {Schaye}, J., {Bower}, R.~G., {et~al.} 2015, \mnras, 450, 1937

\bibitem[{{De Lucia} \& {Blaizot}(2007)}]{LB2007}
{De Lucia}, G. \& {Blaizot}, J. 2007, \mnras, 375, 2

\bibitem[{{de Ravel} {et~al.}(2009){de Ravel}, {Le F{\`e}vre}, {Tresse},
  {Bottini}, {Garilli}, {Le Brun}, {Maccagni}, {Scaramella}, {Scodeggio},
  {Vettolani}, {Zanichelli}, {Adami}, {Arnouts}, {Bardelli}, {Bolzonella},
  {Cappi}, {Charlot}, {Ciliegi}, {Contini}, {Foucaud}, {Franzetti},
  {Gavignaud}, {Guzzo}, {Ilbert}, {Iovino}, {Lamareille}, {McCracken},
  {Marano}, {Marinoni}, {Mazure}, {Meneux}, {Merighi}, {Paltani}, {Pell{\`o}},
  {Pollo}, {Pozzetti}, {Radovich}, {Vergani}, {Zamorani}, {Zucca}, {Bondi},
  {Bongiorno}, {Brinchmann}, {Cucciati}, {de La Torre}, {Gregorini}, {Memeo},
  {Perez-Montero}, {Mellier}, {Merluzzi}, \& {Temporin}}]{RFT2009}
{de Ravel}, L., {Le F{\`e}vre}, O., {Tresse}, L., {et~al.} 2009, \aap, 498, 379

\bibitem[{{Dubois} {et~al.}(2014){Dubois}, {Pichon}, {Welker}, {Le Borgne},
  {Devriendt}, {Laigle}, {Codis}, {Pogosyan}, {Arnouts}, {Benabed}, {Bertin},
  {Blaizot}, {Bouchet}, {Cardoso}, {Colombi}, {de Lapparent}, {Desjacques},
  {Gavazzi}, {Kassin}, {Kimm}, {McCracken}, {Milliard}, {Peirani}, {Prunet},
  {Rouberol}, {Silk}, {Slyz}, {Sousbie}, {Teyssier}, {Tresse}, {Treyer},
  {Vibert}, \& {Volonteri}}]{DPW2014}
{Dubois}, Y., {Pichon}, C., {Welker}, C., {et~al.} 2014, \mnras, 444, 1453

\bibitem[{{Erb} {et~al.}(2014){Erb}, {Steidel}, {Trainor}, {Bogosavljevi{\'c}},
  {Shapley}, {Nestor}, {Kulas}, {Law}, {Strom}, {Rudie}, {Reddy}, {Pettini},
  {Konidaris}, {Mace}, {Matthews}, \& {McLean}}]{EST2014}
{Erb}, D.~K., {Steidel}, C.~C., {Trainor}, R.~F., {et~al.} 2014, \apj, 795, 33

\bibitem[{{Ferreras} {et~al.}(2014){Ferreras}, {Trujillo},
  {M{\'a}rmol-Queralt{\'o}}, {P{\'e}rez-Gonz{\'a}lez}, {Cava}, {Barro},
  {Cenarro}, {Hern{\'a}n-Caballero}, {Cardiel},
  {Rodr{\'{\i}}guez-Zaur{\'{\i}}n}, \& {Cebri{\'a}n}}]{FTM2014}
{Ferreras}, I., {Trujillo}, I., {M{\'a}rmol-Queralt{\'o}}, E., {et~al.} 2014,
  \mnras, 444, 906

\bibitem[{{Genel} {et~al.}(2009){Genel}, {Genzel}, {Bouch{\'e}}, {Naab}, \&
  {Sternberg}}]{GGB2009}
{Genel}, S., {Genzel}, R., {Bouch{\'e}}, N., {Naab}, T., \& {Sternberg}, A.
  2009, \apj, 701, 2002

\bibitem[{{Genel} {et~al.}(2008){Genel}, {Genzel}, {Bouch{\'e}}, {Sternberg},
  {Naab}, {F{\"o}rster Schreiber}, {Shapiro}, {Tacconi}, {Lutz}, {Cresci},
  {Buschkamp}, {Davies}, \& {Hicks}}]{GGB2008}
{Genel}, S., {Genzel}, R., {Bouch{\'e}}, N., {et~al.} 2008, \apj, 688, 789

\bibitem[{{Guo} \& {White}(2008)}]{GW2008}
{Guo}, Q. \& {White}, S.~D.~M. 2008, \mnras, 384, 2

\bibitem[{{Harris} \& {Zaritsky}(2009)}]{HZ2009}
{Harris}, J. \& {Zaritsky}, D. 2009, \aj, 138, 1243

\bibitem[{{Hashimoto} {et~al.}(2013){Hashimoto}, {Ouchi}, {Shimasaku}, {Ono},
  {Nakajima}, {Rauch}, {Lee}, \& {Okamura}}]{HOS2013}
{Hashimoto}, T., {Ouchi}, M., {Shimasaku}, K., {et~al.} 2013, \apj, 765, 70

\bibitem[{{Inami} \& {et al.}(2017)}]{I2017}
{Inami}, H. \& {et al.} 2017, \aap, submitted

\bibitem[{{Jian} {et~al.}(2012){Jian}, {Lin}, \& {Chiueh}}]{JLC2012}
{Jian}, H.-Y., {Lin}, L., \& {Chiueh}, T. 2012, \apj, 754, 26

\bibitem[{{Kampczyk} {et~al.}(2007){Kampczyk}, {Lilly}, {Carollo}, {Scarlata},
  {Feldmann}, {Koekemoer}, {Leauthaud}, {Sargent}, {Taniguchi}, \&
  {Capak}}]{KLC2007}
{Kampczyk}, P., {Lilly}, S.~J., {Carollo}, C.~M., {et~al.} 2007, \apjs, 172,
  329

\bibitem[{{Kartaltepe} {et~al.}(2007){Kartaltepe}, {Sanders}, {Scoville},
  {Calzetti}, {Capak}, {Koekemoer}, {Mobasher}, {Murayama}, {Salvato},
  {Sasaki}, \& {Taniguchi}}]{KSS2007}
{Kartaltepe}, J.~S., {Sanders}, D.~B., {Scoville}, N.~Z., {et~al.} 2007, \apjs,
  172, 320

\bibitem[{{Kaviraj} {et~al.}(2015){Kaviraj}, {Devriendt}, {Dubois}, {Slyz},
  {Welker}, {Pichon}, {Peirani}, \& {Le Borgne}}]{KDD2015}
{Kaviraj}, S., {Devriendt}, J., {Dubois}, Y., {et~al.} 2015, \mnras, 452, 2845

\bibitem[{{Keenan} {et~al.}(2014){Keenan}, {Foucaud}, {De Propris}, {Hsieh},
  {Lin}, {Chou}, {Huang}, {Lin}, \& {Chang}}]{KFP2014}
{Keenan}, R.~C., {Foucaud}, S., {De Propris}, R., {et~al.} 2014, \apj, 795, 157

\bibitem[{{Kere{\v s}} {et~al.}(2005){Kere{\v s}}, {Katz}, {Weinberg}, \&
  {Dav{\'e}}}]{KKW2005}
{Kere{\v s}}, D., {Katz}, N., {Weinberg}, D.~H., \& {Dav{\'e}}, R. 2005,
  \mnras, 363, 2

\bibitem[{{Kitzbichler} \& {White}(2008)}]{KW2008}
{Kitzbichler}, M.~G. \& {White}, S.~D.~M. 2008, \mnras, 391, 1489

\bibitem[{{Koch} {et~al.}(2015){Koch}, {Frank}, {Pasquali}, {Rich}, \&
  {Rabitz}}]{KFP2015}
{Koch}, A., {Frank}, M.~J., {Pasquali}, A., {Rich}, R.~M., \& {Rabitz}, A.
  2015, \apj, 815, 105

\bibitem[{{Kriek} {et~al.}(2009){Kriek}, {van Dokkum}, {Labb{\'e}}, {Franx},
  {Illingworth}, {Marchesini}, \& {Quadri}}]{KVL2009}
{Kriek}, M., {van Dokkum}, P.~G., {Labb{\'e}}, I., {et~al.} 2009, \apj, 700,
  221

\bibitem[{{Labb{\'e}} {et~al.}(2015){Labb{\'e}}, {Oesch}, {Illingworth}, {van
  Dokkum}, {Bouwens}, {Franx}, {Carollo}, {Trenti}, {Holden}, {Smit},
  {Gonz{\'a}lez}, {Magee}, {Stiavelli}, \& {Stefanon}}]{LOI2015}
{Labb{\'e}}, I., {Oesch}, P.~A., {Illingworth}, G.~D., {et~al.} 2015, \apjs,
  221, 23

\bibitem[{{Lagos} {et~al.}(2017){Lagos}, {Stevens}, {Bower}, {Davis},
  {Contreras}, {Padilla}, {Obreschkow}, {Croton}, {Trayford}, {Welker}, \&
  {Theuns}}]{LSB2017}
{Lagos}, C.~d.~P., {Stevens}, A.~R.~H., {Bower}, R.~G., {et~al.} 2017, ArXiv
  e-prints [\eprint[arXiv]{1701.04407}]

\bibitem[{{Le F{\`e}vre} {et~al.}(2000){Le F{\`e}vre}, {Abraham}, {Lilly},
  {Ellis}, {Brinchmann}, {Schade}, {Tresse}, {Colless}, {Crampton},
  {Glazebrook}, {Hammer}, \& {Broadhurst}}]{FAL2000}
{Le F{\`e}vre}, O., {Abraham}, R., {Lilly}, S.~J., {et~al.} 2000, \mnras, 311,
  565

\bibitem[{{Lin} {et~al.}(2004){Lin}, {Koo}, {Willmer}, {Patton}, {Conselice},
  {Yan}, {Coil}, {Cooper}, {Davis}, {Faber}, {Gerke}, {Guhathakurta}, \&
  {Newman}}]{LKW2004}
{Lin}, L., {Koo}, D.~C., {Willmer}, C.~N.~A., {et~al.} 2004, \apjl, 617, L9

\bibitem[{{Lin} {et~al.}(2008){Lin}, {Patton}, {Koo}, {Casteels}, {Conselice},
  {Faber}, {Lotz}, {Willmer}, {Hsieh}, {Chiueh}, {Newman}, {Novak}, {Weiner},
  \& {Cooper}}]{LPK2008}
{Lin}, L., {Patton}, D.~R., {Koo}, D.~C., {et~al.} 2008, \apj, 681, 232

\bibitem[{{L{\'o}pez-Sanjuan} {et~al.}(2009{\natexlab{a}}){L{\'o}pez-Sanjuan},
  {Balcells}, {Garc{\'{\i}}a-Dab{\'o}}, {Prieto}, {Crist{\'o}bal-Hornillos},
  {Eliche-Moral}, {Abreu}, {Erwin}, \& {Guzm{\'a}n}}]{LBG2009b}
{L{\'o}pez-Sanjuan}, C., {Balcells}, M., {Garc{\'{\i}}a-Dab{\'o}}, C.~E.,
  {et~al.} 2009{\natexlab{a}}, \apj

\bibitem[{{L{\'o}pez-Sanjuan} {et~al.}(2009{\natexlab{b}}){L{\'o}pez-Sanjuan},
  {Balcells}, {P{\'e}rez-Gonz{\'a}lez}, {Barro}, {Garc{\'{\i}}a-Dab{\'o}},
  {Gallego}, \& {Zamorano}}]{LBP2009a}
{L{\'o}pez-Sanjuan}, C., {Balcells}, M., {P{\'e}rez-Gonz{\'a}lez}, P.~G.,
  {et~al.} 2009{\natexlab{b}}, \aap

\bibitem[{{L{\'o}pez-Sanjuan} {et~al.}(2015){L{\'o}pez-Sanjuan}, {Cenarro},
  {Varela}, {Viironen}, {Molino}, {Ben{\'{\i}}tez}, {Arnalte-Mur}, {Ascaso},
  {D{\'{\i}}az-Garc{\'{\i}}a}, {Fern{\'a}ndez-Soto}, {Jim{\'e}nez-Teja},
  {M{\'a}rquez}, {Masegosa}, {Moles}, {Povi{\'c}}, {Aguerri}, {Alfaro},
  {Aparicio-Villegas}, {Broadhurst}, {Cabrera-Ca{\~n}o}, {Castander}, {Cepa},
  {Cervi{\~n}o}, {Crist{\'o}bal-Hornillos}, {Del Olmo}, {Gonz{\'a}lez Delgado},
  {Husillos}, {Infante}, {Mart{\'{\i}}nez}, {Perea}, {Prada}, \&
  {Quintana}}]{LCV2015}
{L{\'o}pez-Sanjuan}, C., {Cenarro}, A.~J., {Varela}, J., {et~al.} 2015, \aap,
  576, A53

\bibitem[{{L{\'o}pez-Sanjuan} {et~al.}(2011){L{\'o}pez-Sanjuan}, {Le
  F{\`e}vre}, {de Ravel}, {Cucciati}, {Ilbert}, {Tresse}, {Bardelli},
  {Bolzonella}, {Contini}, {Garilli}, {Guzzo}, {Maccagni}, {McCracken},
  {Mellier}, {Pollo}, {Vergani}, \& {Zucca}}]{2011A&A...530A..20L}
{L{\'o}pez-Sanjuan}, C., {Le F{\`e}vre}, O., {de Ravel}, L., {et~al.} 2011,
  \aap, 530, A20

\bibitem[{{L{\'o}pez-Sanjuan} {et~al.}(2012){L{\'o}pez-Sanjuan}, {Le
  F{\`e}vre}, {Ilbert}, {Tasca}, {Bridge}, {Cucciati}, {Kampczyk}, {Pozzetti},
  {Xu}, {Carollo}, {Contini}, {Kneib}, {Lilly}, {Mainieri}, {Renzini},
  {Sanders}, {Scodeggio}, {Scoville}, {Taniguchi}, {Zamorani}, {Aussel},
  {Bardelli}, {Bolzonella}, {Bongiorno}, {Capak}, {Caputi}, {de la Torre}, {de
  Ravel}, {Franzetti}, {Garilli}, {Iovino}, {Knobel}, {Kova{\v c}},
  {Lamareille}, {Le Borgne}, {Le Brun}, {Le Floc'h}, {Maier}, {McCracken},
  {Mignoli}, {Pell{\'o}}, {Peng}, {P{\'e}rez-Montero}, {Presotto},
  {Ricciardelli}, {Salvato}, {Silverman}, {Tanaka}, {Tresse}, {Vergani},
  {Zucca}, {Barnes}, {Bordoloi}, {Cappi}, {Cimatti}, {Coppa}, {Koekemoer},
  {Liu}, {Moresco}, {Nair}, {Oesch}, {Schawinski}, \& {Welikala}}]{LFI2012}
{L{\'o}pez-Sanjuan}, C., {Le F{\`e}vre}, O., {Ilbert}, O., {et~al.} 2012, \aap,
  548, A7

\bibitem[{{L{\'o}pez-Sanjuan} {et~al.}(2013){L{\'o}pez-Sanjuan}, {Le
  F{\`e}vre}, {Tasca}, {Epinat}, {Amram}, {Contini}, {Garilli},
  {Kissler-Patig}, {Moultaka}, {Paioro}, {Perret}, {Queyrel}, {Tresse},
  {Vergani}, \& {Divoy}}]{LFT2013}
{L{\'o}pez-Sanjuan}, C., {Le F{\`e}vre}, O., {Tasca}, L.~A.~M., {et~al.} 2013,
  \aap, 553, A78

\bibitem[{{Lotz} {et~al.}(2008){Lotz}, {Davis}, {Faber}, {Guhathakurta},
  {Gwyn}, {Huang}, {Koo}, {Le Floc'h}, {Lin}, {Newman}, {Noeske}, {Papovich},
  {Willmer}, {Coil}, {Conselice}, {Cooper}, {Hopkins}, {Metevier}, {Primack},
  {Rieke}, \& {Weiner}}]{LDF2008}
{Lotz}, J.~M., {Davis}, M., {Faber}, S.~M., {et~al.} 2008, \apj, 672, 177

\bibitem[{{Lotz} {et~al.}(2011){Lotz}, {Jonsson}, {Cox}, {Croton}, {Primack},
  {Somerville}, \& {Stewart}}]{LJC2011}
{Lotz}, J.~M., {Jonsson}, P., {Cox}, T.~J., {et~al.} 2011, \apj, 742, 103

\bibitem[{{Lotz} {et~al.}(2010){Lotz}, {Jonsson}, {Cox}, \&
  {Primack}}]{LJP2010}
{Lotz}, J.~M., {Jonsson}, P., {Cox}, T.~J., \& {Primack}, J.~R. 2010, \mnras,
  404, 575

\bibitem[{{Madau} \& {Dickinson}(2014)}]{MD2014}
{Madau}, P. \& {Dickinson}, M. 2014, \araa, 52, 415

\bibitem[{{Maller} {et~al.}(2006){Maller}, {Katz}, {Kere{\v s}}, {Dav{\'e}}, \&
  {Weinberg}}]{MKK2006}
{Maller}, A.~H., {Katz}, N., {Kere{\v s}}, D., {Dav{\'e}}, R., \& {Weinberg},
  D.~H. 2006, \apj, 647, 763

\bibitem[{{Man} {et~al.}(2012){Man}, {Toft}, {Zirm}, {Wuyts}, \& {van der
  Wel}}]{MTZ2012}
{Man}, A.~W.~S., {Toft}, S., {Zirm}, A.~W., {Wuyts}, S., \& {van der Wel}, A.
  2012, \apj, 744, 85

\bibitem[{{Man} {et~al.}(2016){Man}, {Zirm}, \& {Toft}}]{MZT2016}
{Man}, A.~W.~S., {Zirm}, A.~W., \& {Toft}, S. 2016, \apj, 830, 89

\bibitem[{{Maseda} \& {et al.}(2017)}]{M2017}
{Maseda}, M. \& {et al.} 2017, \aap, submitted

\bibitem[{{McLinden} {et~al.}(2011){McLinden}, {Finkelstein}, {Rhoads},
  {Malhotra}, {Hibon}, {Richardson}, {Cresci}, {Quirrenbach}, {Pasquali},
  {Bian}, {Fan}, \& {Woodward}}]{MFR2011}
{McLinden}, E.~M., {Finkelstein}, S.~L., {Rhoads}, J.~E., {et~al.} 2011, \apj,
  730, 136

\bibitem[{{Mihos} \& {Hernquist}(1994)}]{MH1994}
{Mihos}, J.~C. \& {Hernquist}, L. 1994, \apjl, 437, L47

\bibitem[{{Mihos} \& {Hernquist}(1996)}]{MH1996}
{Mihos}, J.~C. \& {Hernquist}, L. 1996, \apj, 464, 641

\bibitem[{{Moreno} {et~al.}(2013){Moreno}, {Bluck}, {Ellison}, {Patton},
  {Torrey}, \& {Moster}}]{MBE2013}
{Moreno}, J., {Bluck}, A.~F.~L., {Ellison}, S.~L., {et~al.} 2013, \mnras, 436,
  1765

\bibitem[{{Moster} {et~al.}(2011){Moster}, {Somerville}, {Newman}, \&
  {Rix}}]{MSN2011}
{Moster}, B.~P., {Somerville}, R.~S., {Newman}, J.~A., \& {Rix}, H.-W. 2011,
  \apj, 731, 113

\bibitem[{{Naab} \& {Burkert}(2003)}]{NB2003}
{Naab}, T. \& {Burkert}, A. 2003, \apj, 597, 893

\bibitem[{{Nikolic} {et~al.}(2004){Nikolic}, {Cullen}, \&
  {Alexander}}]{NCA2004}
{Nikolic}, B., {Cullen}, H., \& {Alexander}, P. 2004, \mnras, 355, 874

\bibitem[{{Ocvirk} {et~al.}(2008){Ocvirk}, {Pichon}, \& {Teyssier}}]{OPT2008}
{Ocvirk}, P., {Pichon}, C., \& {Teyssier}, R. 2008, \mnras, 390, 1326

\bibitem[{{Patton} \& {Atfield}(2008)}]{PA2008}
{Patton}, D.~R. \& {Atfield}, J.~E. 2008, \apj, 685, 235

\bibitem[{{Patton} {et~al.}(2000){Patton}, {Carlberg}, {Marzke}, {Pritchet},
  {da Costa}, \& {Pellegrini}}]{PCM2000}
{Patton}, D.~R., {Carlberg}, R.~G., {Marzke}, R.~O., {et~al.} 2000, \apj, 536,
  153

\bibitem[{{Patton} {et~al.}(1997){Patton}, {Pritchet}, {Yee}, {Ellingson}, \&
  {Carlberg}}]{PPY1997}
{Patton}, D.~R., {Pritchet}, C.~J., {Yee}, H.~K.~C., {Ellingson}, E., \&
  {Carlberg}, R.~G. 1997, \apj, 475, 29

\bibitem[{{Perez} {et~al.}(2006){Perez}, {Tissera}, {Scannapieco}, {Lambas}, \&
  {de Rossi}}]{PTS2006}
{Perez}, M.~J., {Tissera}, P.~B., {Scannapieco}, C., {Lambas}, D.~G., \& {de
  Rossi}, M.~E. 2006, \aap, 459, 361

\bibitem[{{Perret} {et~al.}(2014){Perret}, {Renaud}, {Epinat}, {Amram},
  {Bournaud}, {Contini}, {Teyssier}, \& {Lambert}}]{PRE2014}
{Perret}, V., {Renaud}, F., {Epinat}, B., {et~al.} 2014, \aap, 562, A1

\bibitem[{{Popesso} {et~al.}(2009){Popesso}, {Dickinson}, {Nonino}, {Vanzella},
  {Daddi}, {Fosbury}, {Kuntschner}, {Mainieri}, {Cristiani}, {Cesarsky},
  {Giavalisco}, {Renzini}, \& {GOODS Team}}]{PDN2009}
{Popesso}, P., {Dickinson}, M., {Nonino}, M., {et~al.} 2009, \aap, 494, 443

\bibitem[{{Qu} {et~al.}(2017){Qu}, {Helly}, {Bower}, {Theuns}, {Crain},
  {Frenk}, {Furlong}, {McAlpine}, {Schaller}, {Schaye}, \& {White}}]{QHB2017}
{Qu}, Y., {Helly}, J.~C., {Bower}, R.~G., {et~al.} 2017, \mnras, 464, 1659

\bibitem[{{Rafelski} {et~al.}(2015){Rafelski}, {Teplitz}, {Gardner}, {Coe},
  {Bond}, {Koekemoer}, {Grogin}, {Kurczynski}, {McGrath}, {Bourque}, {Atek},
  {Brown}, {Colbert}, {Codoreanu}, {Ferguson}, {Finkelstein}, {Gawiser},
  {Giavalisco}, {Gronwall}, {Hanish}, {Lee}, {Mehta}, {de Mello},
  {Ravindranath}, {Ryan}, {Scarlata}, {Siana}, {Soto}, \& {Voyer}}]{RTG2015}
{Rafelski}, M., {Teplitz}, H.~I., {Gardner}, J.~P., {et~al.} 2015, \aj, 150, 31

\bibitem[{{Rodriguez-Gomez} {et~al.}(2015){Rodriguez-Gomez}, {Genel},
  {Vogelsberger}, {Sijacki}, {Pillepich}, {Sales}, {Torrey}, {Snyder},
  {Nelson}, {Springel}, {Ma}, \& {Hernquist}}]{RGV2015}
{Rodriguez-Gomez}, V., {Genel}, S., {Vogelsberger}, M., {et~al.} 2015, \mnras,
  449, 49

\bibitem[{{Rodriguez-Gomez} {et~al.}(2016){Rodriguez-Gomez}, {Pillepich},
  {Sales}, {Genel}, {Vogelsberger}, {Zhu}, {Wellons}, {Nelson}, {Torrey},
  {Springel}, {Ma}, \& {Hernquist}}]{RPS2016}
{Rodriguez-Gomez}, V., {Pillepich}, A., {Sales}, L.~V., {et~al.} 2016, \mnras,
  458, 2371

\bibitem[{{Ryan} {et~al.}(2008){Ryan}, {Cohen}, {Windhorst}, \&
  {Silk}}]{RCW2008}
{Ryan}, Jr., R.~E., {Cohen}, S.~H., {Windhorst}, R.~A., \& {Silk}, J. 2008,
  \apj, 678, 751

\bibitem[{{Schaye} {et~al.}(2015){Schaye}, {Crain}, {Bower}, {Furlong},
  {Schaller}, {Theuns}, {Dalla Vecchia}, {Frenk}, {McCarthy}, {Helly},
  {Jenkins}, {Rosas-Guevara}, {White}, {Baes}, {Booth}, {Camps}, {Navarro},
  {Qu}, {Rahmati}, {Sawala}, {Thomas}, \& {Trayford}}]{SCB2015}
{Schaye}, J., {Crain}, R.~A., {Bower}, R.~G., {et~al.} 2015, \mnras, 446, 521

\bibitem[{{Shibuya} {et~al.}(2014){Shibuya}, {Ouchi}, {Nakajima}, {Hashimoto},
  {Ono}, {Rauch}, {Gauthier}, {Shimasaku}, {Goto}, {Mori}, \&
  {Umemura.}}]{SON2014}
{Shibuya}, T., {Ouchi}, M., {Nakajima}, K., {et~al.} 2014, \apj, 788, 74

\bibitem[{{Snyder} {et~al.}(2017){Snyder}, {Lotz}, {Rodriguez-Gomez},
  {Guimar{\~a}es}, {Torrey}, \& {Hernquist}}]{SLR2017}
{Snyder}, G.~F., {Lotz}, J.~M., {Rodriguez-Gomez}, V., {et~al.} 2017, \mnras,
  468, 207

\bibitem[{{Somerville} {et~al.}(2001){Somerville}, {Primack}, \&
  {Faber}}]{SPF2001}
{Somerville}, R.~S., {Primack}, J.~R., \& {Faber}, S.~M. 2001, \mnras, 320, 504

\bibitem[{{Stewart} {et~al.}(2009){Stewart}, {Bullock}, {Barton}, \&
  {Wechsler}}]{SBB2009}
{Stewart}, K.~R., {Bullock}, J.~S., {Barton}, E.~J., \& {Wechsler}, R.~H. 2009,
  \apj, 702, 1005

\bibitem[{{Tasca} {et~al.}(2014){Tasca}, {Le F{\`e}vre}, {L{\'o}pez-Sanjuan},
  {Wang}, {Cassata}, {Garilli}, {Ilbert}, {Le Brun}, {Lemaux}, {Maccagni},
  {Tresse}, {Bardelli}, {Contini}, {Charlot}, {Cucciati}, {Fontana},
  {Giavalisco}, {Kneib}, {Salvato}, {Taniguchi}, {Vergani}, {Zamorani}, \&
  {Zucca}}]{TFL2014}
{Tasca}, L.~A.~M., {Le F{\`e}vre}, O., {L{\'o}pez-Sanjuan}, C., {et~al.} 2014,
  \aap, 565, A10

\bibitem[{{Verhamme} \& {et al.}(2017)}]{V2017}
{Verhamme}, A. \& {et al.} 2017, \aap, submitted

\bibitem[{{Verhamme} {et~al.}(2015){Verhamme}, {Orlitov{\'a}}, {Schaerer}, \&
  {Hayes}}]{V2015}
{Verhamme}, A., {Orlitov{\'a}}, I., {Schaerer}, D., \& {Hayes}, M. 2015, \aap,
  578, A7

\bibitem[{{Vogelsberger} {et~al.}(2014){Vogelsberger}, {Genel}, {Springel},
  {Torrey}, {Sijacki}, {Xu}, {Snyder}, {Nelson}, \& {Hernquist}}]{VGS2014}
{Vogelsberger}, M., {Genel}, S., {Springel}, V., {et~al.} 2014, \mnras, 444,
  1518

\bibitem[{{White} \& {Rees}(1978)}]{WR1978}
{White}, S.~D.~M. \& {Rees}, M.~J. 1978, \mnras, 183, 341

\bibitem[{{Williams} {et~al.}(2000){Williams}, {Baum}, {Bergeron}, {Bernstein},
  {Blacker}, {Boyle}, {Brown}, {Carollo}, {Casertano}, {Covarrubias}, {de
  Mello}, {Dickinson}, {Espey}, {Ferguson}, {Fruchter}, {Gardner}, {Gonnella},
  {Hayes}, {Hewett}, {Heyer}, {Hook}, {Irwin}, {Jones}, {Kaiser}, {Levay},
  {Lubenow}, {Lucas}, {Mack}, {MacKenty}, {Madau}, {Makidon}, {Martin},
  {Mazzuca}, {Mutchler}, {Norris}, {Perriello}, {Phillips}, {Postman}, {Royle},
  {Sahu}, {Savaglio}, {Sherwin}, {Smith}, {Stiavelli}, {Suntzeff}, {Teplitz},
  {van der Marel}, {Walker}, {Weymann}, {Wiggs}, {Williger}, {Wilson},
  {Zacharias}, \& {Zurek}}]{WBB2000}
{Williams}, R.~E., {Baum}, S., {Bergeron}, L.~E., {et~al.} 2000, \aj, 120, 2735

\bibitem[{{Xu} {et~al.}(2012){Xu}, {Zhao}, {Scoville}, {Capak}, {Drory}, \&
  {Gao}}]{XSB2012}
{Xu}, C.~K., {Zhao}, Y., {Scoville}, N., {et~al.} 2012, \apj, 747, 85

\end{thebibliography}

\longtab{
\begin{longtable}{r c c r r c c r r r c}
\caption{\label{table:1} Basic properties for the sample of major galaxy close pairs in the HDF-S, \textsf{udf-10,} and UDF-Mosaic. Labels 1 and 2 denote the primary and secondary galaxy, respectively. Cols.\,(1) and (5): Identification number in the MUSE-based catalogues of Bacon et al.\,(2015) for HDF-S galaxies, and Inami et al.\,(2017) for HUDF galaxies. Cols.\,(2) and (6): MUSE spectroscopic redshift with associated confidence level (2 and 3 $=$ secure redshift, 1 $=$ possible redshift, see Inami et al.\,2017 for details) in cols.\,(3) and (7). Cols.\,(4) and (8): Stellar masses in logarithmic units. Cols.\,(9) and (10): Projected separation (in kpc) and velocity difference (in \kms) between the two galaxies in the pair, respectively.}\\
\hline\hline                 
&&&&&&&&&&\\
MUSE ID$_1$ & $z_1$ & zconf$_1$ & M$^{\star}_1$ & MUSE ID$_2$ & $z_2$ & zconf$_2$ & M$^{\star}_2$ & r$_p$ & $\Delta_v$ & MUSE field \\    
$-$ & $-$ & $-$ & [log(\Msun)] & $-$ & $-$ & $-$ & [log(\Msun)] & [kpc] & [\kms] & $-$ \\
(1) & (2) & (3) & (4) & (5) & (6) & (7) & (8) & (9) & (10) & (11)\\ &&&&&&&&&&\\
\\
\hline                        
&&&&&&&&&&\\
 29 & 0.831 & 3 & 10.44  & 58 & 0.832 & 1 & 10.21 & 25.3 & 138 & HDF-S \\
 45 & 1.155 & 3 & 9.90  & 134 & 1.155 & 2 & 9.78 & 20.0 & 56 & HDF-S \\ 
 50 & 2.672 & 3 & 10.96  & 55 & 2.674 & 3 & 10.78 & 6.6 & 119 & HDF-S \\ 
 88 & 1.360 & 2 & 8.70  & 589 & 1.359 & 2 & 8.08 & 5.0 & 15& HDF-S \\
 183 & 3.374 & 2 & 9.81  & 261 & 3.375 & 1 & 9.81 & 2.4 & 78 & HDF-S \\ 
 433 & 3.470 & 2 & 7.35  & 478 & 3.469 & 1 & 7.17 & 20.9 & 145 & HDF-S \\ 
 441 & 4.695 & 2 & 7.85  & 453 & 4.701 & 1 & 7.49 & 24.6 & 438 & HDF-S \\ 
 492 & 5.760 & 2 & 8.22  & 577 & 5.764 & 1 & 8.51 & 18.6 & 16 & HDF-S \\ 
 551 & 3.180 & 2 & 9.81  & 578 & 3.180 & 1 & 9.81 & 3.8 & 59 & HDF-S \\ 
 &&&&&&&&&&\\
\hline  
 &&&&&&&&&&\\
3  & 0.622 & 3  & 9.92 & 9  & 0.619 & 3  & 10.23 & 14.6 & 411  & \textsf{udf-10} \\ 
24  & 2.544 & 3  & 9.75 & 35  & 2.543 & 3  & 10.04 & 14.5 & 62  & \textsf{udf-10} \\ 
30  & 1.096 & 3  & 8.94 & 84  & 1.096 & 3  & 8.81 & 35.7 & 54  & \textsf{udf-10} \\ 
32  & 1.307 & 3  & 9.23 & 77  & 1.310 & 1  & 8.68 & 33.8 & 413  & \textsf{udf-10} \\ 
32  & 1.307 & 3  & 9.23 & 121  & 1.306 & 3  & 8.56 & 11.7 & 72  & \textsf{udf-10} \\ 
46  & 1.413 & 3  & 9.31 & 92  & 1.414 & 3  & 8.54 & 8.2 & 21  & \textsf{udf-10} \\ 
61  & 2.454 & 3  & 9.58 & 67  & 2.449 & 3  & 10.18 & 12.2 & 399  & \textsf{udf-10} \\ 
65  & 1.307 & 3  & 8.97 & 77  & 1.310 & 1  & 8.68 & 13.6 & 378  & \textsf{udf-10} \\ 
96  & 0.622 & 3  & 7.69 & 108  & 0.622 & 3  & 7.78 & 20.7 & 63  & \textsf{udf-10} \\ 
344  & 3.471 & 2  & 8.52 & 6871  & 3.474 & 1  & 9.15 & 19.7 & 195  & \textsf{udf-10} \\ 
399  & 5.137 & 2  & 7.52 & 627  & 5.135 & 2  & 7.15 & 26.2 & 99  & \textsf{udf-10} \\ 
399  & 5.137 & 2  & 7.52 & 6339  & 5.131 & 2  & 6.98 & 27.8 & 305  & \textsf{udf-10} \\ 
430  & 4.514 & 2  & 8.64 & 6340  & 4.510 & 2  & 8.97 & 30.5 & 223  & \textsf{udf-10} \\ 
430  & 4.514 & 2  & 8.64 & 6342  & 4.514 & 2  & 8.52 & 4.0 & 3  & \textsf{udf-10} \\ 
627  & 5.135 & 2  & 7.15 & 6339  & 5.131 & 2  & 6.98 & 22.8 & 206  & \textsf{udf-10} \\ 
6302  & 3.473 & 2  & 9.18 & 6925  & 3.474 & 2  & 9.63 & 32.7 & 68  & \textsf{udf-10} \\ 
 &&&&&&&&&&\\
\hline  
 &&&&&&&&&&\\
430  & 4.513 & 2  & 7.84 & 7197  & 4.513 & 2  & 8.18 & 30.8 & 2  & UDF-Mosaic \\ 
891  & 0.227 & 3  & 7.84 & 6891  & 0.227 & 3  & 7.15 & 21.2 & 35  & UDF-Mosaic \\ 
899  & 1.097 & 3  & 10.18 & 934  & 1.096 & 3  & 9.79 & 30.5 & 94  & UDF-Mosaic \\ 
950  & 0.993 & 3  & 9.00 & 1107  & 0.993 & 3  & 8.73 & 8.3 & 3  & UDF-Mosaic \\ 
997  & 1.041 & 3  & 8.93 & 1454  & 1.041 & 3  & 8.69 & 32.6 & 24  & UDF-Mosaic \\ 
999  & 1.608 & 3  & 9.93 & 1268  & 1.609 & 2  & 9.71 & 7.4 & 46  & UDF-Mosaic \\ 
1027  & 0.219 & 3  & 7.63 & 1167  & 0.219 & 3  & 7.08 & 16.5 & 43  & UDF-Mosaic \\ 
1044  & 2.028 & 3  & 10.17 & 1048  & 2.028 & 2  & 10.08 & 31.8 & 81  & UDF-Mosaic \\ 
1065  & 0.522 & 3  & 8.21 & 1444  & 0.523 & 3  & 7.61 & 28.1 & 290  & UDF-Mosaic \\ 
1178  & 2.691 & 3  & 9.69 & 1279  & 2.691 & 1  & 9.66 & 32.5 & 65  & UDF-Mosaic \\ 
1188  & 1.412 & 2  & 9.61 & 1219  & 1.413 & 2  & 9.12 & 28.0 & 118  & UDF-Mosaic \\ 
1267  & 1.866 & 3  & 9.58 & 6947  & 1.866 & 2  & 9.76 & 32.5 & 10  & UDF-Mosaic \\ 
1341  & 1.413 & 3  & 9.12 & 1373  & 1.413 & 3  & 8.89 & 9.3 & 36  & UDF-Mosaic \\ 
1345  & 1.095 & 3  & 8.57 & 1605  & 1.095 & 3  & 8.71 & 26.9 & 37  & UDF-Mosaic \\ 
1545  & 0.992 & 3  & 8.33 & 6991  & 0.991 & 3  & 8.26 & 19.0 & 156  & UDF-Mosaic \\ 
1561  & 0.733 & 3  & 7.68 & 1644  & 0.732 & 3  & 7.52 & 7.0 & 67  & UDF-Mosaic \\ 
1611  & 0.666 & 3  & 7.79 & 1688  & 0.665 & 1  & 7.27 & 22.9 & 150  & UDF-Mosaic \\ 
1678  & 1.425 & 2  & 8.76 & 7101  & 1.427 & 2  & 8.67 & 32.2 & 262  & UDF-Mosaic \\ 
1990  & 1.219 & 3  & 8.55 & 6885  & 1.216 & 2  & 8.94 & 20.4 & 496  & UDF-Mosaic \\ 
2071  & 4.930 & 2  & 9.33 & 6412  & 4.928 & 2  & 9.37 & 14.0 & 98  & UDF-Mosaic \\ 
2672  & 3.439 & 2  & 8.78 & 7351  & 3.433 & 2  & 8.03 & 33.2 & 400  & UDF-Mosaic \\ 
2695  & 3.067 & 2  & 7.66 & 3430  & 3.061 & 2  & 7.40 & 30.5 & 436  & UDF-Mosaic \\ 
2757  & 5.380 & 2  & 7.91 & 5398  & 5.382 & 1  & 7.22 & 33.4 & 86  & UDF-Mosaic \\ 
3840  & 4.813 & 2  & 7.30 & 5508  & 4.807 & 2  & 6.89 & 27.2 & 318  & UDF-Mosaic \\ 
4532  & 3.438 & 2  & 8.52 & 7221  & 3.435 & 2  & 8.54 & 34.1 & 215  & UDF-Mosaic \\ 
4542  & 4.811 & 2  & 7.16 & 5882  & 4.811 & 2  & 6.85 & 26.2 & 2  & UDF-Mosaic \\ 
6402  & 4.372 & 2  & 8.40 & 7311  & 4.372 & 2  & 8.47 & 20.7 & 6  & UDF-Mosaic \\ 
6517  & 3.432 & 2  & 8.86 & 6531  & 3.432 & 1  & 8.58 & 28.6 & 3  & UDF-Mosaic \\ 
6923  & 3.433 & 2  & 7.53 & 7283  & 3.432 & 2  & 8.00 & 21.2 & 62  & UDF-Mosaic \\ 
7285  & 5.486 & 2  & 7.74 & 7353  & 5.485 & 2  & 7.48 & 33.8 & 46  & UDF-Mosaic \\ 
 &&&&&&&&&&\\

\hline                                   
\end{longtable}
}
\end{document}